\documentclass[journal]{IEEEtran}

\ifCLASSINFOpdf
   \usepackage[pdftex]{graphicx}
   \usepackage{epstopdf}
   \graphicspath{{../pdf/}{../jpeg/}}
   \DeclareGraphicsExtensions{.pdf,.jpeg,.png}
\else
   \usepackage[dvips]{graphicx}
   \graphicspath{{../eps/}}
\fi

\usepackage{amsmath}
\usepackage{amssymb}
\usepackage{cite}
\usepackage{graphicx}
\usepackage{algorithm}
\usepackage{algpseudocode}
\usepackage{subfigure}
\usepackage{multirow} 
\usepackage{longtable}
\usepackage{threeparttable}
\usepackage{caption3}
\usepackage{array}
\usepackage{cases}
\usepackage{color}
\usepackage[T1]{fontenc}
\usepackage[utf8]{inputenc}
\usepackage{authblk}
\DeclareMathOperator{\Tr}{Tr}
\DeclareMathOperator*{\argmax}{argmax}

\usepackage{algorithm}
\usepackage{algpseudocode}

\begin{document}
\title{Multiuser Wirelessly Powered Backscatter Communications: Nonlinearity, Waveform Design and SINR-Energy Tradeoff}

\author{Zati~Bayani~Zawawi,
        Yang~Huang
        and~Bruno~Clerckx
\thanks{Z. B. Zawawi and B. Clerckx are with the Department of Electrical and Electronic Engineering, Imperial College London, South Kensington Campus London SW72AZ, United Kingdom (email: \{z.zawawimohd-zawawi13, b.clerckx\}@imperial.ac.uk). Y. Huang is with the Department of Information and Communication Engineering, Nanjing University of Aeronautics and Astronautics, Nanjing 211100, China (email: yang.huang.ceie@nuaa.edu.cn). This work has been partially supported by the EPSRC of the UK under grant EP/P003885/1.}
}

%



\maketitle

\begin{abstract}
Wireless power transfer and backscatter communications have emerged as promising solutions for energizing and communicating with power limited devices. Despite some progress in wirelessly powered backscatter communications, the focus has been concentrated on backscatter and energy harvester. Recently, significant progress has been made on the design of transmit multisine waveform, adaptive to the Channel State Information at the transmitter (CSIT), in point-to-point backscatter system. In this paper, we leverage the work and study the design of transmit multisine waveform in a multiuser backscatter system, made of one transmitter, one reader and multiple tags active simultaneously. We derive an efficient algorithm to optimize the transmit waveform so as to identify the tradeoff between the amount of energy harvested at the tags and the reliability of the communication, measured in terms of Signal-to-Interference-plus-Noise Ratio (SINR) at the reader. Performance with the optimized waveform based on linear and nonlinear energy harvester (EH) models are studied. Numerical results demonstrate the benefits of accounting for the energy harvester nonlinearity, multiuser diversity, frequency diversity and multisine waveform adaptive to the CSIT to enlarge the SINR-energy region.
\end{abstract}


\section{Introduction}
In recent years, the study of Wireless Information and Power Transfer (WIPT) has attracted significant interest among researchers \cite{ClerckxFundamental}. WIPT has appeared in multiple forms such as simultaneous wireless information and power transfer (SWIPT) where power and information are simultaneously transmit to the receivers \cite{Archi_SWIPT}, wireless powered communication where the transmitter transmits power to the receiver, which is then converted to DC power and reused for information transmission \cite{Hju2014} and wireless powered backscatter communication where the transmitter transmits power/unmodulated signal to the backscatter and the backscatter harvests the energy from the received signal or backscatter the received signal back to the reader \cite{Gyang2015}.

Wirelessly powered backscatter communication is a promising technology for low-power communication systems because tags can operate with low power as they do not require RF components to generate carrier signal. Although backscatter communications has originally been limited to simple radio frequency identification (RFID) applications, it has received a renewed interests in recent years with advances in backscatter communication theory, including coding \cite{Boyer}, beamforming \cite{Gyang2015}, performance analysis of large networks \cite{Han, Bacha}.

RF transmitter in backscatter communications typically transmit sinusoidal continuous waveform (CW). Significant progress has recently been made on the waveform design in wireless power transfer (WPT), in order to improve the efficiency and DC output power of the energy harvester. In particular, \cite{Boaventura2011} studies the nonlinear behaviour of RF-to-DC converter and suggests that significant DC power gain can be obtained by using multisine waveform with zero phase between the sinewaves. In \cite{M.S.Trotter2009}, a multisine waveform transmission has been introduced to extend the reading signal. \cite{M.S.Trotter2010} presented multisine waveform method to improve power sensitivity of the tags and conduct a survey of reading range improvement for several commercial RFID tags. Motivated by the promising gains, a systematic approach towards waveform design for WPT was proposed in \cite{Clerckx2016}. In \cite{Clerckx2016}, optimized multisine waveforms, adaptive to the CSIT, have been shown to provide significant gain by exploiting the rectifier nonlinearity and the frequency selectivity of the channel.

Backscatter communication can leverage this recent progress in WPT waveform design and depart from the conventional CW transmission. Recently, \cite{Clerckx2017} studies the tradeoff between harvested energy and backscatter communication in point-to-point deployment. It is noted that waveform design for WPT and backscatter communication is different since backscatter communication is subject to a tradeoff between harvested energy at the tag and SNR at the reader \cite{Clerckx2017}. In \cite{Clerckx2017}, by assuming that the CSIT is available, a systematic design of multisine transmit waveform is derived to enlarge the tradeoff region, therefore boosting the overall performance of backscatter communication.

In this paper, we leverage the work in \cite{Clerckx2017} and study the design of transmit multisine waveform in a multiuser backscatter system. To the best of the authors' knowledge, waveform design in a multiuser wirelessly powered backscatter system has not been addressed yet. In this paper, the nonlinear energy harvesting model introduced in \cite{Clerckx2017, Clerckx2016} is used to model the energy harvester \footnote{In the literature, there are different types of energy harvesting models. The energy harvesting model in \cite{Clerckx2017, Clerckx2016} relates the output DC current/power to the input signal through the diode I-V characteristics in the nonlinear region. \cite{Archi_SWIPT} has introduced the linear model of the energy harvester, which ignores the diode nonlinearity. In contrast to the nonlinear model in \cite{Clerckx2017, Clerckx2016}, the linear model may not accurately demonstrate the nonlinearity of the rectifier. Another type of model is the saturation nonlinear model proposed in \cite{saturation} that models the saturation of the output DC power due to the rectifier operating in diode breakdown region. The difference between the linear model, the diode nonlinear model (used in this paper) and the saturation nonlinear model has been discussed in length in \cite{ClerckxFundamental, swipt_nonlin}. The diode nonlinear model is more suited for multiple reasons. First, backscatter communications operate in the low power regime (typical input power to rectifier is of the order of -20dBm). Second, the diode nonlinear model does reflect the dependence on the input signal power and shape and can be used for waveform design (in contrast with the saturation model that cannot be used for such purpose since it is fitted to a given pre-defined signal). Third, the diode nonlinearity is beneficial to system performance in the low power-regime and therefore should be exploited in backscatter communications (in contrast with the saturation nonlinearity that is detrimental to performance and therefore should be avoided by proper rectifier design). The readers are referred to \cite{ClerckxFundamental} and Remark 5 in \cite{swipt_nonlin} for extensive discussions on the various linear and nonlinear models.}. This is a general model of the diode nonlinearity, that is fundamentally motivated by the physics of the rectifier and is applicable to various rectifier topologies \cite{ClerckxLowComplexity}. The model and the resulting signal design has been validated through circuit simulations in \cite{Clerckx2016, ClerckxLowComplexity} and experimentations \cite{KimPrototyping}. The model also relies on a Taylor expansion of the diode characteristics, as commonly done in the RF literature \cite{Boaventura2011}. The contributions of this paper are summarized in the following paragraphs.

{\it First}, by making use of the nonlinear rectenna model introduced in \cite{Clerckx2016}, we design multisine transmit waveform and characterize the SINR-energy region in a $K$-tags wirelessly powered backscatter system. In contrast to \cite{Clerckx2016} that only consider transmit waveform for power transfer, this paper needs to consider transmit waveform for power and backscatter communication simultaneously. In \cite{Clerckx2017}, the tradeoff region for a point-to-point wireless powered communication system has been studied. Here, the optimal phases of multisine waveform weights can be obtained in a closed-form. For given optimal phases, optimizing the magnitudes of the waveform results in maximizing non-convex posynomial and can be solved by using reversed Geometric Programming (GP). Note that the algorithm proposed in \cite{Clerckx2017} for $K=1$ cannot be used for a general K tags system. For a general $K$-tags system, optimizations over the phases and the magnitudes cannot be decoupled \cite{Huang2016, YH}, such that a closed form solution cannot be obtained and therefore the reversed GP method cannot be applied. Therefore, a new optimization algorithm is proposed in this paper to jointly optimize the phases and the magnitudes of the multisine waveform for the $K$-tags system, by making use of the matrix formulation and optimization technique developed in \cite{Huang2016, YH}. Additionally, in contrast to the design for a pure Wireless Power Transfer (WPT) system (as in \cite{Huang2016, YH}), the design in this paper needs to optimize the receive combiners at the reader, such that the SINR with respect to a tag can be higher than a threshold. The optimization of the transmit waveform and the receive combiners are coupled. Therefore, an iterative algorithm is proposed to jointly optimize the transmit waveform and the receive combiner iteratively.

{\it Second}, the performance of waveform design based on linear and nonlinear rectenna models is studied. It is observed that waveform design based on a nonlinear model (4th order truncation) has a larger SINR-energy region than those obtained based on the linear model. Hence, the rectenna nonlinearity, if properly exploited in the waveform design, is beneficial to the backscatter communication system performance.

{\it Third}, several interesting observations have been found through numerical simulation. Power allocation across multiple sinewaves has been found to enlarge the tradeoff region by exploiting the frequency selectivity of the channel and the nonlinearity of the rectifier. The gain achieved by waveforms optimized based on the nonlinear model over those designed based on linear model increases as the number of sinewaves increases. It is also observed that the waveform design benefits from multiuser diversity to enlarge the tradeoff between SINR and total amount of energy harvested at the tags. In addition, the performance gap between waveform optimized based on nonlinear and linear models increases as the number of tag increases. Nevertheless, the average amount of energy harvested at each tag decreases as the number of tag increases. This is because, the waveform has to be designed to meet the SINR requirement at all tags, and therefore decrease the average amount of energy harvested at each tag. Another observation is that waveform adaptive to the CSIT is beneficial to maximize the harvested energy for given SINR constraint, such that the tradeoff region is enlarged.

The remainder of this paper is organized as follows. In section II, we discuss the system model. In section III, we discuss waveform optimization for multiuser backscatter communications. We present the simulation results in section IV and conclude the paper in section V.

\ In this paper, a bold capital letter and a bold lower case letter represents a matrix and a vector, respectively. The notations $(.)^*$, $(.)^{\star}$, $(.)^T$, $(.)^H$, $\Tr(.)$, $\left\vert{}.\right\vert{}$ and $\left\Vert{}\bf .\right\Vert{}$ represent the conjugate, optimal solution, transpose, conjugate transpose, trace, absolute value and 2-norm, respectively. $\mathcal{A}\{.\}$ indicates the DC component of a signal and $\Re\{.\}$ refers to the real number. The notation $0\leq x \perp y \geq 0$ denotes that $x \geq 0$, $y \geq 0$ and $x.y=0$.

\section{System Model}
\begin{figure}[!t]
\centering
\hspace{-0.1in}
\includegraphics[width=3in]{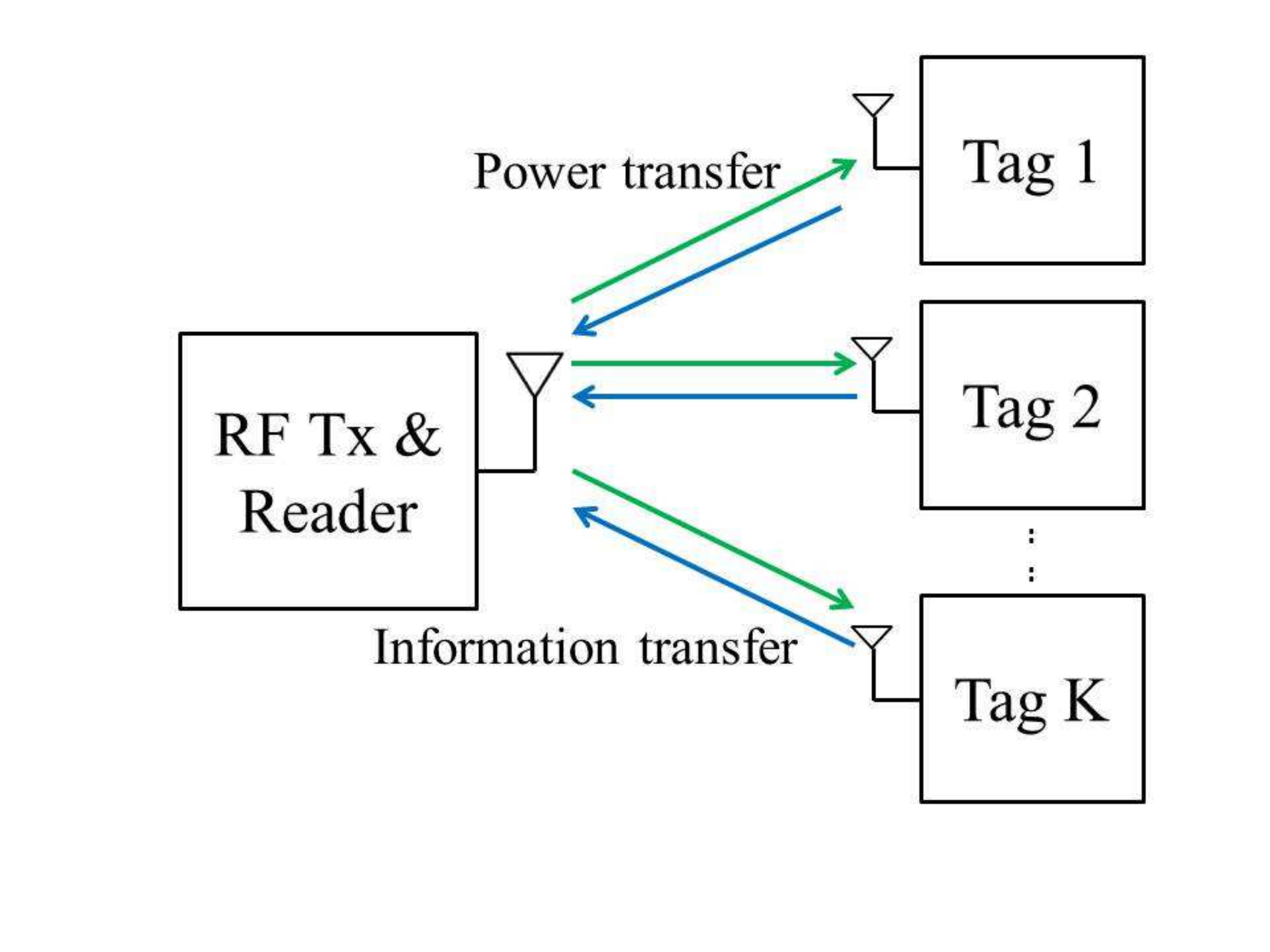}
\caption{Wirelessly powered backscatter communication with multiple tags.}
\label{system}
\end{figure}

In this section, we provide a system model for multiuser wirelessly powered backscatter communication as shown in Fig. \ref{system}. An RF transmitter transmits a multisine waveform, with $N$ sinewaves to $K$ tags. Each tag converts the incoming RF signal into DC current and transfers information to the reader (which is co-located with the RF transmitter) by backscattering modulation. RF transmitter/reader and each tag have a single antenna. We assume that the RF transmitter has perfect knowledge of the forward channel and the backward channel (i.e., channel from tag to reader).

\subsection{The Transmit Waveform}
The transmitter transmits the deterministic multisine waveform given by
\begin{equation}
x(t)=\sum_{n=0}^{N-1}s_{n}\cos(\omega_{n}t+\phi_{n})=\Re\{\sum_{n=0}^{N-1}{x_{n}e^{i\omega_nt}}\},
\end{equation}
where $x_{n}=s_{n}e^{i\phi_{n}}$. Here, $s_{n}$ and $\phi_{n}$ are the amplitude and the phase of the $n$-th sinewave at frequency $\omega_n$, respectively. The magnitudes and phases in (1) can be collected into vectors ${\bf s}$ and ${\bf \Phi}$. The transmit waveform is subject to transmit power constraint $\mathcal{E}\{{\vert{} x\vert{}}^2\}=\frac{1}{2}{\Vert{}{\bf s}\Vert{}}^2\leq P$.

After multipath propagation, the received waveform at the $j$-th tag can be written as
\begin{equation}
\begin{aligned}
y_{j}(t)&=\sum_{n=0}^{N-1}s_{n}A_{j,n}\cos(\omega_nt+\psi_{j,n})=\Re\{\sum_{n=0}^{N-1}{h_{j,n}x_{n}e^{i\omega_nt}}\},
\end{aligned}
\end{equation}
where $h_{j,n}=A_{j,n}e^{i\bar{\psi}_{j,n}}=\sum_{l=0}^{L-1}\alpha_{j,l}e^{i(-\omega_n\tau_{j,l}+\varepsilon_{j,l})}$ is the frequency response of the channel between the transmitter and the $j$-th tag at frequency component $n$ and $\psi_{j,n}={\bar \psi}_{j,n}+\phi_{n}$. $\tau_{j,l}$, $\varepsilon_{j,l}$ and $\alpha_{j,l}$ are the delay, phase and amplitude of the l-th path from the transmitter to the $j$-th tag, respectively.

\subsection{The Tag Operation}
We consider that each tag employs a simple binary modulation to transfer information to the reader as in \cite{Clerckx2017}. Binary 0 refers to a perfect impedance matching that completely absorbs the incoming signal and binary 1 refers to a perfect impedance mismatch that completely reflects the incoming signal. The signal absorbed during binary 0 operation is conveyed to the rectifier for energy harvesting, while the signal reflected during binary 1 operation is backscattered to the reader. The reader performs information detection of sequence bit 0 and 1 from the incoming backscattered signal.

\subsection{The Energy Harvester}
We will use the same rectenna model as in \cite{Clerckx2017, Clerckx2016}. The rectenna is made of an antenna and a rectifier. 
Received power is transferred from the antenna to the rectifier through the matching network. We assume a lossless antenna model with voltage source $v_{s,j}(t)$ at the $j$-th tag and antenna impedance $R_{ant}=50 \Omega$ is connected to an input impedance of the rectifier and matching network $Z_{in,j}$. With perfect impedance matching \footnote{The challenge in designing the waveform is to construct an analytical rectenna model that is accurate and tractable enough to be able to optimize the transmit waveform. In this paper, perfect impedance matching across all frequencies is assumed to balance the complexity and accuracy of the model. Note that in practice, perfect impedance matching for multisine transmission cannot be achieved at all the frequencies. Despite this, the waveform design approach that assumes perfect matching has been validated by circuit simulations in \cite{Clerckx2016} and \cite{ClerckxLowComplexity}, where the matching networks used in circuit simulations are designed by exploiting ADS harmonic balance simulation. It is also to be noted that in this paper, the inter-frequency spacing (of the order of MHz at most) is very small compared to the carrier frequency (GHz), making the transmission narrowband from an RF design perspective. Even though the impedance mismatch may occur in a multisine transmission, the impact is minimal in a narrowband multisine transmission.} where $Z_{in,j}=R_{ant}$ during binary 0 operation, all incoming RF power is completely transferred to the rectifier such that $P_{av,j}=\mathcal{E}\{{\vert{} y_j(t)\vert{}}^2\}=\mathcal{E}\{{\vert{} v_{in,j}(t)\vert{}}^2\}/R_{ant}$.

\begin{figure}[!t]
\centering
\includegraphics[width=2.5in]{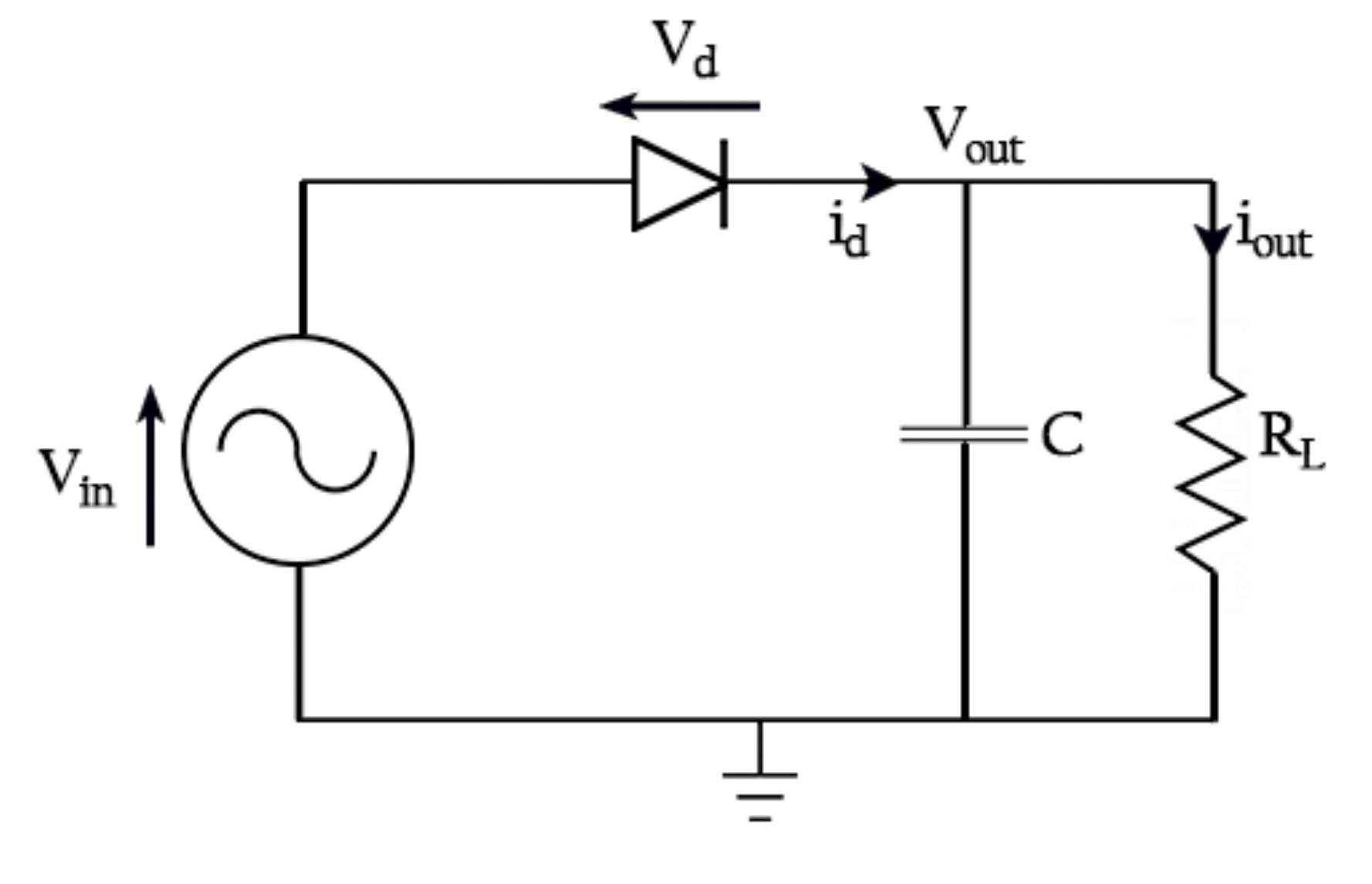}
\caption{A single diode rectifier circuit.}
\label{rectifier circuit}
\end{figure}

Consider a simple rectifier circuit made of a nonlinear diode followed by a low pass filter and a load (i.e. $R_L$) as shown in Fig. \ref{rectifier circuit}. The current flowing through the diode at the $j$-th tag is given as $i_{d,j}(t)=i_s(e^{\frac{v_{d,j}(t)}{nv_t}-1}-1)$ where $v_{d,j}(t)=v_{in,j}(t)-v_{out,j}(t)$ is the voltage drop across the diode, $i_s$ is the reverse bias saturation current, $v_t$ is the thermal voltage and $n$ is the ideality factor. By using Taylor expansion of the exponential function around a fixed voltage drop $v_{d,j}=a_j$, the diode current can be written as
\begin{equation}
i_{d,j}(t)=\sum_{u=0}^{\infty}k'_{u,j}{(v_{d,j}(t)-a_j)}^u
\end{equation}
where $k'_{0,j}=i_s(e^{\frac{a_j}{nv_t}}-1)$, $k'_{u,j}=i_s(\frac{e^{\frac{a_j}{nv_t}}}{u!{(nv_t)}^u})$ for $u=1,..,\infty$. As stated in \cite{Clerckx2016}, this energy harvesting model is valid only for small signal where the diode works in the nonlinear region of diode I-V characteristic. When the signal becomes large, the diode series resistance dominates diode behaviour and the diode will be driven into the linear region of I-V characteristic. In this case, the Taylor series based model and the assumptions made does not hold. For more discussion on the energy harvesting model, the readers are referred to Section III-B in \cite{Clerckx2016}.

Assuming a steady state response and ideal rectifier, the output voltage delivered to the load is a constant $v_{out,j}$. When $a_j=\mathcal{E}\{v_{d,j}(t)\}=-v_{out,j}$, (3) can be written as
\begin{equation} 
i_{d,j}(t)=\sum_{u=0}^{\infty}k'_{u,j}{v_{in,j}(t)}^u=\sum_{u=0}^{\infty}k'_{u,j}R_{ant}^{u/2}{y_j(t)}^u.
\end{equation}
By ignoring the terms higher than 4 \footnote{4th order has been described in \cite{Clerckx2016, Boaventura2011} to be the minimum order to demonstrate the nonlinearity of the rectification.}, the DC component in (4) can be written as 
\begin{equation} 
i_{d,j}(t)=\sum_{u=0}^{4}k'_{u,j}R_{ant}^{u/2}\mathcal{A}\{{y_j(t)}^u\}.
\end{equation}
Following \cite{Clerckx2016}, the maximization of (5) is equivalent to maximizing 
\begin{equation}
z_{DC,j}=k_2R_{ant}\mathcal{A}\{{y_j(t)}^2\}+k_4R_{ant}^{2}\mathcal{A}\{{y_j(t)}^4\},
\end{equation}
where $k_u=\frac{i_s}{u!{(nv_t)}^u}$, 
\begin{equation}
\mathcal{A}\{y_{j}(t)^2\}=\frac{1}{2}\sum_{n=1}^{N}s_{n}^2A_{j,n}^2
\end{equation}
and
\begin{equation}
\begin{aligned}
&\mathcal{A}\{y_{j}(t)^4\}=\\ &\frac{3}{8}\Big[\sum_{\substack{n_1, n_2, n_3, n_4\\n_1+n_2\\=n_3+n_4}}[s_{n_j}A_{j,n_m}]\cos(\psi_{j,n_1}+\psi_{j,n_2}-\psi_{j,n_3}-\psi_{j,n_4})\Big].
\end{aligned}
\end{equation}
For $i_s=5\mu A$, $n=1.05$ and $v_t=25.86m$V, typical values of $k_2=0.0034$ and $k_4=0.3829$ \footnote{These values will be used in any evaluation throughout this paper.}. In the sequel, (6) is called the nonlinear model of the energy harvester. The linear model of the energy harvester is obtained as a special case by ignoring the 4th order term. By defining the weight \footnote{The weights are introduced to provide fairness and priorities among tags. The weights can be adjusted over time.} for energy harvested at tag $j$ as $c_j\geq 0$, we are interested in a general metric consisting of a weigthed sum of $z_{DC,j}$ given by $Z_{DC}=\sum_{j=1}^Kc_jz_{DC,j}$.

\subsection{The Backscattered Signal}
The tag completely absorbs and reflects the incoming signal for binary information 0 and 1, respectively. The received signal at the reader is given as
\begin{equation}
\begin{aligned}
z(t)&=\Re\{\sum_{n=1}^N \sum_{j=1}^K m_j h_{j,n}h_{j,n}^bx_{n}e^{i2\pi f_nt}\}+v(t)
\end{aligned}
\end{equation}
where $m_j=0,1$ is the binary information for $j$-th tag, $h_{j,n}^b=A_{j,n}^be^{i\bar{\psi}_{j,n}^b}$ is the frequency response of the backward channel between the $j$-th tag and the reader at frequency component $n$ and $v(t)$ is AWGN. After applying a product detector to each frequency component and assuming perfect low pass filtering, the baseband received signal at subcarrier $n$ is given by
\begin{equation}
z_{n}=\sum_{j=1}^K h_{j,n}h_{j,n}^bx_{n}m_j+v_{n}.
\end{equation}
$v_{n}$ is the complex white Gaussian noise at frequency component $n$ with variance $\sigma^2$. Let ${\bf z}={[z_{1}, .., z_{N}]}^T$, the signal ${\tilde z}_j$ used for information detection of the $j$-th tag can be written as
\begin{equation}
{\tilde z}_j={\bf g}_j^H{\bf z},
\end{equation}
where ${\bf g}_j^H \in \mathbb{C}^{1 \times N}$ is the receive combiner at the reader.

It is noted that the SINR of each tag is invariant to the scaling of ${\bf g}_j$. Thus, without loss of generality, we define $\Vert{\bf g}_j\Vert=1$. The SINR for information detection of the $j$-th tag can be obtained by
\begin{equation}
\begin{aligned}
\rho_j({\bf w}, {\bf g}_{j})&=\frac{{\vert {\bf g}_{j}^{H} {\bf H}_j{\bf w}\vert}^2}{\sigma^2+{\vert {\bf g}_{j}^{H}  {\bf \tilde H}_j {\bf w}\vert}^2},
\end{aligned}
\end{equation}
where ${\bf w}={[x_1,..,x_N]}^T$, ${\bf H}_j=\mbox{diag}(h_{j,1}{h}_{j,1}^b,..,h_{j,N}{h}_{j,N}^b)$ and ${\bf \tilde H}_j=\mbox{diag}(\sum_{u\not= j}^K  h_{u,1} h_{u,1}^b,..,\sum_{u\not= j}^K  h_{u,N} h_{u,N}^b)$.

\subsection{CSIT Assumption}
We assume that the RF transmitter has a perfect knowledge of CSI of the forward channel $h_{j,n}$ $\forall j, n$ and the backscatter channel $h_{j,n}h_{j,n}^b$ $\forall j, n$, so that the transmit waveform can be shaped as a function of the channel states to maximize $Z_{DC}$ and $\rho_j$. The backscatter channel gain $h_{j,n}h_{j,n}^b$ is equal to the product of the forward channel gain $h_{j,n}$ and the backward channel gain $h_{j,n}^b$. To estimate the backscatter channel, a pilot signal can be delivered by the transmitter. This pilot signal goes through the backscatter channel and is received by the reader. Then, by performing least-square estimation \cite{Hassibi2003}, the backscatter channel can be estimated. In the presence of channel reciprocity, $h_{j,n} = h_{j,n}^b = \sqrt{h_{j,n}h_{j,n}^b}$. On the other hand, for $h_{j,n} \neq h_{j,n}^b$, an additional pilot signal can be delivered by the tag, such that the backward channel can be estimated at the reader. Thereby, the forward channel gain $h_{j,n}$ can be easily computed. A long pilot transmission phase can improve the accuracy of channel estimation, as well as add to the overhead. Thus, the length of the pilot transmission phase can be optimized so as to balance the estimation accuracy and the overhead \cite{Hassibi2003, Mishra2018}. It is worth noting that this paper focus on the design of multisine transmit waveform and the effect of rectenna nonlinearity in a backscatter communication system. The problem of optimizing the length of the pilot transmission phase can be studied in the future.

We also assume that the reader has a perfect knowledge of CSI of the backscatter channel $h_{j,n}h_{j,n}^b$ $\forall j, n$, so that the receive combining can be performed to maximize $\rho_j$.

\section{Waveform Optimizaton and SINR-Energy Tradeoff Characterization}
In this section, we discuss waveform optimization and characterize the tradeoff between achievable SINR and energy harvested for a multiuser backscatter system .
The optimization problem is formulated as maximizing $Z_{DC}$ given that an SINR constraint at each tag and an average transmit power constraint are satisfied. The optimization problem is given by
\begin {equation}
\underset{{\bf w}, {\bf g}_{j}}{\max}\{Z_{DC}: \rho_j\geq {\bar \rho}_j, {\Vert{\bf w}\Vert}^2\leq 2P, \Vert{\bf g}_j\Vert=1, \forall j\},
\end {equation}
where ${\bar \rho}_j$ is the SINR constraint for the $j$-th tag and $P$ is the transmit power constraint. Note that for certain channel realizations and under the transmit power constraint, there may not be any ${\bf w}$ and ${\bf g}_j$ satisfying the SINR constraints. Therefore, we first check the feasibility of problem (13) as discussed in the following section. 

\subsection{Feasibility Problem}
For a given transmit power constraint and channel realization, problem (13) is feasible when the SINR constraint ${\bar \rho}_j$ for all tags can be satisfied. In the case that the problem has no solution, (13) is infeasible. Note that the waveform is only optimized when (13) is a feasible problem.

The coupling of the optimization variables ${\bf w}$ and ${\bf g}_j$ in the SINR constraints makes problem (13) non-convex. In this paper, we propose an algorithm based on alternating optimization \cite{Alter} that optimizes ${\bf w}$ and ${\bf g}_j$ iteratively. As a result, we formulate the optimization problem as maximizing an auxiliary variable $\delta=\underset{j=1,..,K}{\min} \{ {\rho}_j/{\bar \rho}_j\}$, which is used to check if all SINR constraints are satisfied. Specifically, the feasibility problem is formulated as
\begin {equation}
\begin{aligned}
&\underset{{\bf w}, {\bf g}_j}{\max} &&\delta\\
&\mbox{s. t.} &&\frac{{\vert {\bf g}_{j}^{H} {\bf H}_j{\bf w}\vert}^2}{\sigma^2{\Vert {\bf g}_{j}\Vert}^2+{\vert {\bf g}_{j}^{H}  {\bf \tilde H}_j {\bf w}\vert}^2} \geq \delta {\bar \rho}_j, &&&\forall j \\
& &&{\Vert{\bf w}\Vert}^2\leq 2P,\ \ \ \ \\\
& &&\Vert{\bf g}_j\Vert=1. &&&\forall j
\end{aligned}
\end {equation}
Here, (14) is feasible if $\delta^\star \geq 1$, which implies that $\rho_j\geq {\bar \rho}_j$ for all tags (i.e., the achievable SINR at all tags are larger than their corresponding SINR constraint). During the alternating optimization, $\delta$ is updated by using bisection search method as in \cite{CLi2016} over $\delta_{min}\leq\delta\leq\delta_{max}$. $\delta_{min}$ and $\delta_{max}$ are chosen such that $\delta_{min}<1$ and $\delta_{max}>1$.  

For a given ${\bf w}^{}$, the optimal receive combiner that maximizes ${\rho}_j$ in (14) is given in \cite{Palomar2003} as
\begin{equation}
\begin{aligned}
{\bf g}_j^{\star}=&\frac{{({\bf \tilde H}_j{\bf w}{\bf w}^{H}{\bf \tilde H}_j^{H}+{\bf I}\sigma^2)}^{-1}{\bf H}_j{\bf w}}{\Vert {({\bf \tilde H}_j{\bf w}{\bf w}^{H}{\bf \tilde H}_j^{H}+{\bf I}\sigma^2)}^{-1}{\bf H}_j{\bf w}\Vert}.
\end{aligned}
\end{equation}\\

For a given ${\bf g}_j$, as shown in \cite{JQiu2011, Bengtsson99}, we can adjust the direction of ${\bf w}$ such that ${\bf g}_{j}^{H} {\bf H}_j{\bf w}$ is real and non-negative without affecting the value of ${\vert{\bf g}_{j}^{H} {\bf H}_j{\bf w}\vert}$ \footnote{Here, we assume that $N\geq K$ so that we can always get a ${\bf w}$ such that ${\bf g}_{j}^{H} {\bf H}_j{\bf w}= {\vert{\bf g}_{j}^{H} {\bf H}_j{\bf w}\vert}$ for all $K$ tags.}. Accordingly, (14) can be recast as
\begin {equation}
\begin{aligned}
&\underset{{\bf w}}{\max} &&\delta\\
&\mbox{s. t.} &&{( {\bf g}_{j}^{H} {\bf H}_j{\bf w})}^2 \geq \delta {\bar \rho}_j(\sigma^2{\Vert {\bf g}_{j}\Vert}^2+{\vert {\bf g}_{j}^{H}  {\bf \tilde H}_j {\bf w}\vert}^2),\ \forall j\\
& &&{\Vert{\bf w}_{j}\Vert}^2\leq 2P,\ \ \ \ \ \ \ \ \ \ \ \ \ \ \ \ \ \ \ \ \ \ \ \ \ \ \ \ \ \ \ \ \ \ \ \ \forall j\\
& &&{\bf g}_{j}^{H} {\bf H}_j{\bf w}= \Re{({\bf g}_{j}^{H} {\bf H}_j{\bf w})}.\ \ \ \ \ \ \ \ \ \ \ \ \ \ \ \ \ \ \ \ \ \ \ \forall j\\
\end{aligned}
\end {equation}
It can be found that (16) is a SOCP problem and can be solved by using CVX \cite{CVX}. Specifically, at iteration $l$, $\delta$ is updated such that in case (16) is feasible, then $\delta_{min}^{(l)}=\delta$. Otherwise, $\delta_{max}^{(l)}=\delta$. Note that for a given ${\bf w}^{\star(l-1)}$, (15) provides an optimal ${\bf g}_j^{\star(l)}$ that maximizes ${\rho}_j$. However, it might not satisfy the SINR constraint ${\bar \rho}_j$ (i.e., $\delta^{\star}<1$).  Therefore, ${\bf g}_j$ and (16) are optimize iteratively until convergence or $\delta_{min}\geq 1$. The iterative algorithm is summarized in Algorithm 1.

According to Algorithm 1, for given ${\bf w}^{\star(l-1)}$, ${{\bf g}_j^{\star}}^{(l)}$ is obtained by using (15). Subsequently, for given ${{\bf g}_j^{\star}}^{(l)}$, ${{\bf w}}^{\star(l)}$ is optimized by using (16). Note that $\delta^{\star}=\delta_{min}^{\star(l)}$ (as in line 12). For convenience, we define ${\bf g}={[{{\bf g}_1}^T,.., {{\bf g}_K}^T]}^T$ and the objective in (14) as the objective function $\delta = \delta_{min}({\bf g}, {\bf w})$. Because for given ${\bf w}^{\star(l-1)}$, ${{\bf g}_j^{\star}}^{(l)}$ from (15) is the optimal solution that maximize ${\rho}_j$ and for given ${{\bf g}^{\star}}^{(l)}$, problem (16) is convex, it shows that $\delta_{min}({\bf g}^{\star(l-1)}, {\bf w}^{\star(l-1)})\leq\delta_{min}({\bf g}^{\star(l)}, {\bf w}^{\star(l)})$ \footnote{Since $\delta_{min}^{\star(l)}\geq \delta_{min}^{\star(l-1)}$, we set $\delta_{min}^{(l)}=\delta_{min}^{(l-1)}$ in line 3 of Algorithm 1.}. Hence, we can find that the sequence $\delta_{min}$ monotonically increases. As $\delta$ is upper bounded by $\delta_{max}$, Algorithm 1 always converges. It can be observed that $\delta_{min}({\bf g}^{\star(l-1)}, {\bf w}^{\star(l-1)})\leq\delta_{min}({\bf g}^{\star(l-1)}, {\bf w}^{\star(l)})\leq\delta_{min}({\bf g}^{\star(l)}, {\bf w}^{\star(l)})$. As $l \to \infty$, it can be shown that Algorithm 1 converges to a stationary point of (14) \cite{CLi2016, Razaviyayn2013}.
\begin{algorithm}[t]
\caption{Feasibility Algorithm}
\label{Algthm_A1}
\begin{algorithmic}[1]
\State \textbf{Initialize}: $\delta_{min}^{(0)}=0$, $\delta_{max}^{(0)}>1$, ${\bf w}^{(0)}$, $\epsilon$, $l=0$
\Repeat
	\State $l=l+1$, $\delta_{min}^{(l)}=\delta_{min}^{(l-1)}$ and $\delta_{max}^{(l)}=\delta_{max}^{(0)}$.
    \State Update ${\bf g}_j^{\star(l)}$ by using (15) for $j=1,..,K$.
    \While{$\delta_{max}^{(l)}-\delta_{min}^{(l)}>\epsilon$}
    \State $\delta=(\delta_{max}^{(l)}+\delta_{min}^{(l)})/2$.
    \State If (16) is feasible, $\delta_{min}^{(l)}=\delta$, store ${\bf w}^\star$.
    \State Else $\delta_{max}^{(l)}=\delta$.
    \EndWhile
    \State Update ${\bf w}^{\star(l)}\gets {\bf w}^\star$ and $\delta_{min}^{\star(l)}\gets \delta_{min}^{(l)}$.
\Until{$\delta_{min}^{\star(l)}-\delta_{min}^{\star(l-1)} \leq\epsilon$ or $\delta_{min}^{\star(l)}\geq 1$}.
\State Update ${\delta}^{\star}=\delta_{min}^{\star(l)}$.
\end{algorithmic}
\end{algorithm}
%
\subsection{Waveform Optimization}
In this section, we discuss waveform optimization in a multiuser system as formulated in (13). Recall that in a point-to-point system \cite{Clerckx2017}, exploiting reversed GP to optimize the waveform requires initial choice of phases before the magnitudes are optimized. In order to maximize $Z_{DC}$, waveform phases are chosen such that the $\cos$(.) in (8) is equal to $0$. In a point-to-point system, the optimal phases can be obtained in closed-form \cite{Clerckx2017}. However, closed-form solution cannot be obtained for a system with $K$ tags. Therefore, we propose an algorithm that jointly optimized the phases and the magnitudes, by making use of the matrix formulation introduced in \cite{Huang2016, YH} \footnote{However, this approach cannot be extended to a truncation order higher than 4.}. 

\begin{figure}[!t]
\centering
\hspace{-0.1in}
\includegraphics[width=3.6in]{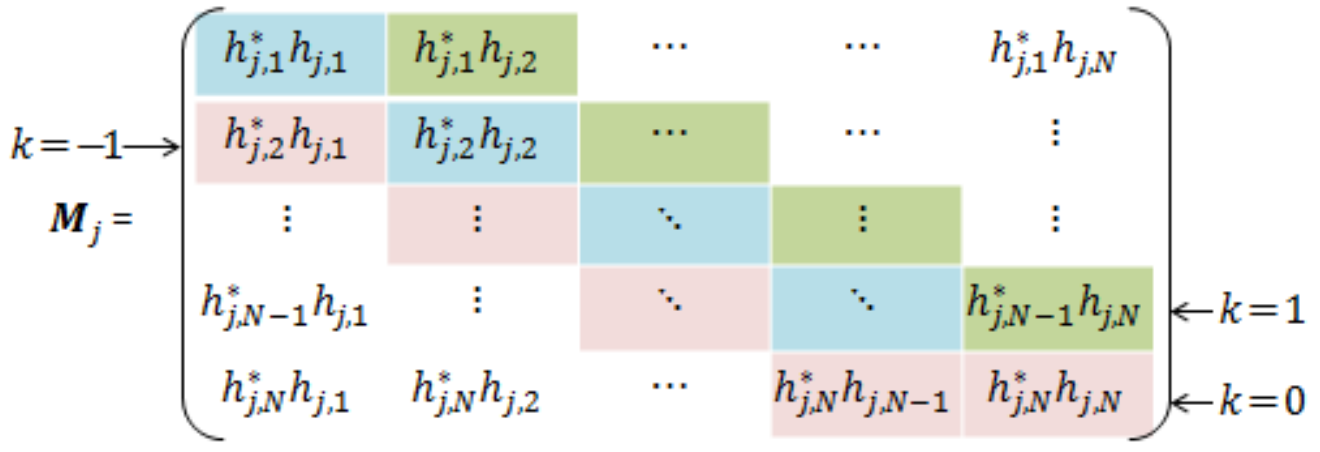}
\caption{An $N$-by-$N$ matrix ${\bf M}_j$.}
\label{Matrix}
\end{figure}
\
By letting ${\bf h}_{j}={[{h}_{j,1},..,{h}_{j,N}]}^T$, we define ${\bf M}_{j}={\bf h}_{j}^*{\bf h}_{j}^T$. As shown in Fig. \ref{Matrix}, $k=0$ is the index of the main diagonal of ${\bf M}_{j}$, $k \in \{ 1,..,N-1\}$ is the index of the k-th diagonal above the main diagonal and $k \in \{ -1,..,-(N-1)\}$ is the index of the $\vert k\vert$-th diagonal below the main diagonal. We define ${\bf M}_{j,k}$ by retaining the k-th diagonal ${\bf M}_{j}$ while setting all other entries as 0. Thus, $Z_{DC}$ can be written as in \eqref{z_DC} where $\beta_2=k_2R_{ant}$ and $\beta_4=k_4R_{ant}^2$.

\par \begin{table*}
\begin{align}\label{z_DC}
Z_{DC}=&\sum_{j=1}^K c_j \Big(\frac{1}{2} \beta_2\sum_n{w}_n^{*}{h}_{j,n}^*{h}_{j,n}{w}_n+\frac{3}{8}\beta_4\sum_{\substack{n_1, n_2, n_3, n_4\\n_1-n_3=-(n_2-n_4)}} {w}_{n_3}^{*}{h}_{j,n_3}^*{h}_{j,n_1}{w}_{n_1}{{w}_{n_4}^{*}{h}_{j,n_4}^*{h}_{j,n_2}{w}_{n_2}}\Big)\nonumber \\
=&\sum_{j=1}^K c_j \Big(\frac{1}{2} \beta_2{\bf w}^H{\bf M}_{j,0}{\bf w}+\frac{3}{8}\beta_4{\bf w}^H{\bf M}_{j,0}{\bf w}{({\bf w}^H{\bf M}_{j,0}{\bf w})}^H+\frac{3}{4}\beta_4\sum_{k=1}^{N-1}{\bf w}^H{\bf M}_{j,k}{\bf w}{({\bf w}^H{\bf M}_{j,k}{\bf w})}^H\Big).
\end{align}\hrulefill
\end{table*}

$Z_{DC}$ in (17) is a quartic polynomial, which in general makes problem (13) NP-hard. In order to address the quartic polynomial problem, auxiliary variables $t_k$ for $k=0,..,N-1$ are introduced such that ${\bf w}^H{\bf M}_{j,k}{\bf w}=t_{j,k}$ \cite{Huang2016, YH}. However, it is noted that for all $j$ with $k \not= 0$, ${\bf M}_{j,k}$ is not a Hermitian matrix. Therefore, the term ${\bf w}^H{\bf M}_{j,k}{\bf w}$ is a bilinear function which may also lead to a NP-hard problem \cite{YH}. Hence, an auxiliary ${\bf X}={\bf w}{\bf w}^H$ is defined as in \cite{YH} to linearize the bilinear term such that we can write ${\bf w}^H{\bf M}_{j,k}{\bf w}=\Tr\{{\bf M}_{j,k}{\bf X}\}=t_{j,k}$. By letting ${\bf A}_0=diag\{-3\beta_4/8,-3\beta_4/4,..,-3\beta_4/4\}\preceq 0$ and ${\bf t}_j={[t_{j,0},..,t_{j,N-1}]}^T$, we can write $q_j({\bf t}_j)\triangleq {\bf t}_j^H{\bf A}_0{\bf t}_j=-3\beta_4/8t_{j,0}t_{j,0}^*-3\beta_4/4\sum_{k=1}^{N-1}t_{j,k}t_{j,k}^*$. Therefore, (17) can be written as

\begin{equation}
\begin{aligned}
Z_{DC}&=\sum_{j=1}^K c_j \big(\frac{\beta_2}{2}t_{j,0}-q_j({\bf t}_j)\big)
\end{aligned}
\end{equation}
and (12) can be written as
\begin{equation}
\begin{aligned}
\rho_j({\bf X}, {\bf g}_{j})&=\frac{\Tr{({\bf H}_j^H{\bf g}_{j}{\bf g}_{j}^{H} {\bf H}_j{\bf X})}}{\sigma^2+\Tr{({\bf \tilde H}_j^H{\bf g}_{j}{\bf g}_{j}^{H} {\bf \tilde H}_j{\bf X})}}.\\
\end{aligned}
\end{equation}

Hence, problem (13) can be recast as
\begin {subequations}
\begin{align}
&\underset{\gamma, {\bf t}_{j},{\bf X}, {\bf g}_{j}}{\min} &&\gamma\\
&\mbox{s. t.} &&\sum_{j=1}^K c_j(-\beta_2t_{j,0}+{q}_j({\bf t}_j))-\gamma \leq 0,\\
& && \frac{\Tr{({\bf H}_j^H{\bf g}_{j}{\bf g}_{j}^{H} {\bf H}_j{\bf X})}}{\sigma^2+\Tr{({\bf \tilde H}_j^H{\bf g}_{j}{\bf g}_{j}^{H} {\bf \tilde H}_j{\bf X})}} \geq {\bar \rho}_j,\ \ \ \ \ \ \ \ \ \forall j\\
& &&\Tr\{{\bf M}_{j,k}{\bf X}\}=t_{j,k},\ \ \ \ \ \ \ \ \ \ \ \ \ \ \ \ \ \ \ \ \forall j,k\\
& &&\Tr\{{\bf M}_{j,k}^H{\bf X}\}=t_{j,k}^*,\ \ \ \ \ \ \ \ \ \ \ \ \ \ \ \forall j,k \not= 0\\
& &&\Tr\{{\bf X}\}\leq 2P,\\
& &&{\mbox {rank}}\{{\bf X}\}=1,\\
& && \Vert{\bf g}_j\Vert=1.\ \ \ \ \ \ \ \ \ \ \ \ \ \ \ \ \ \ \ \ \ \ \ \ \ \ \ \ \ \ \ \ \ \ \ \forall j
\end{align}
\end {subequations}

It can be seen that the non-convex quadratic constraint (20b), the coupled optimization variables in (20c) and the rank constraint (20g) make problem (20) intractable. Meanwhile, the necessary condition for optimizing $\gamma$, ${\bf t}_j$, ${\bf X}$ and ${\bf g}_j$ is nothing but to reduce the value of $\gamma$. Motivated by this necessary condition, we propose an iterative algorithm to solve problem (20). We first relax the rank constraint such that (20) can be written as
\begin {equation}
\begin{aligned}
&\underset{\gamma, {\bf t}_{j},{\bf X}, {\bf g}_{j}}{\min} &&\gamma\\
& && (20b),(20c),(20d),(20e),(20f), (20h).
\end{aligned}
\end {equation}

As in \cite{YH}, we then linearize (20b) to solve (21). Specifically, at iteration $l$, the non-convex term $q_j({\bf t}_j)$ for $j=1,..,K$ in (20b) is approximated at ${\bf t_j}^{(l-1)}$ (the optimal ${\bf t}_j^{\star}$ achieved at iteration $(l-1)$) as a linear function by its first-order Taylor expansion given as ${\tilde q}_j({\bf t}_j,{\bf t}_j^{(l-1)})=2\Re\{{[{\bf t}_j^{(l-1)}]}^H{\bf A}_0{\bf t}_j\}-{[{\bf t}_j^{(l-1)}]}^H{\bf A}_0{\bf t}_j^{(l-1)}$. Thus, at the $l$-th iteration, the following approximate problem is solved
\begin {subequations}
\begin{align}
&\underset{\gamma, {\bf t}_{j},{\bf X}, {\bf g}_{j}}{\min} &&\gamma\\
&\mbox{s. t.} &&\sum_{j=1}^K c_j(-\beta_2t_{j,0}+{\tilde q}_j({\bf t}_j,{\bf t}_j^{(l-1)})-\gamma \leq 0,\\
& && (20c),(20d),(20e),(20f), (20h) \nonumber.
\end{align}
\end {subequations}
Since $-{q}_j({\bf t}_j)$ is convex, it can be shown that ${q}_j({\bf t}_j^{})\leq{\tilde q}_j({\bf t}_j^{},{\bf t}_j^{(l-1)})$. Therefore, ${\tilde q}_j({\bf t}_j^{(l)},{\bf t}_j^{(l)})={q}_j({\bf t}_j^{(l)})\leq{\tilde q}_j({\bf t}_j^{(l)},{\bf t}_j^{(l-1)})$, which indicates that the solution of (22) always satisfy (20b). 

In the following, we show that the value of $\gamma$ decreases over iterations. Unfortunately, (22) is non-convex with respect to ${\bf X}$ and ${\bf g}_j$. Therefore, we optimize ${\bf X}$ and $ {\bf g}_j$ by using alternating algorithm. Specifically at the $l$-th iteration of Successive Convex Approximation (SCA), we first optimize ${\bf g}_j^{(l)}$ for given ${\bf X}^{(l-1)}$. Then ${\bf X}^{(l)}$ is optimized with the updated ${\bf g}_j^{(l)}$. 

For a given ${\bf X}$, problem (22) can be reduced to
\begin {equation}
\begin{aligned}
&\underset{{\bf g}_{j}}{\max} &&\frac{{\bf g}_j^H{\bf H}_j{\bf X}{\bf H}_j^H{\bf g}_j}{{\bf g}_j^H(\sigma^2{\bf I}+{\bf \tilde H}_j{\bf X}{\bf \tilde H}_j^H){\bf g}_j} &&\forall j\\
&\mbox{s. t.} &&\Vert{\bf g}_j\Vert=1.
\end{aligned}
\end {equation}\\
(23) can be transformed to $\rho_j({\bf \tilde g}_j)=\frac{{\bf \tilde g}_j^H{\bf D}{\bf \tilde g}_j}{{\bf \tilde g}_j^H{\bf \tilde g}_j}$ where ${\bf D}={({\bf C})}^{-1}{\bf H}_j{\bf X}{\bf H}_j^H{({\bf C}^H)}^{-1}$, ${\bf C}{\bf C}^H$ is the Cholesky decomposition of $\sigma^2{\bf I}+{\bf \tilde H}_j{\bf X}{\bf \tilde H}_j^H$ and ${\bf \tilde g}_j={\bf C}^H{\bf g}_j$. Here, the optimal ${\bf \tilde g}_j^\star$ is the eigenvector corresponding to the largest eigenvalue of ${\bf D}$. Accordingly, ${\bf g}_j^\star={({\bf C}^H)}^{-1}{\bf \tilde g}_j^\star/{\Vert {({\bf C}^H)}^{-1}{\bf \tilde g}_j^\star\Vert}$.

For given ${\bf g}_j$, problem (22) can be reduced to
\begin {subequations}
\begin{align}
&\underset{\gamma, {\bf t}_{j}, {\bf X}\succeq 0}{\min} &&\gamma\\
&\mbox{s. t.} &&\sum_{j=1}^K c_j(-\beta_2t_{j,0}+{\tilde q}_j({\bf t}_j,{\bf t}_j^{(l-1)}))-\gamma \leq 0,\\
& && \frac{\Tr{({\bf H}_j^H{\bf g}_{j}{\bf g}_{j}^{H} {\bf H}_j{\bf X})}}{\sigma^2{\Vert{\bf g}_{j}\Vert}^2+\Tr{({\bf \tilde H}_j^H{\bf g}_{j}{\bf g}_{j}^{H} {\bf \tilde H}_j{\bf X})}} \geq {\bar \rho}_j,\ \ \forall j\\
& &&\Tr\{{\bf M}_{j,k}{\bf X}\}=t_{j,k},\ \ \ \ \ \ \ \ \ \ \ \ \ \ \ \ \ \ \ \ \forall j,k\\
& &&\Tr\{{\bf M}_{j,k}^H{\bf X}\}=t_{j,k}^*,\ \ \ \ \ \ \ \ \ \ \ \ \ \ \ \ \ \ \ \ \forall j,k\\
& &&\Tr\{{\bf X}\}\leq 2P.
\end{align}
\end {subequations}
(24) is a SDP problem and can be solved by CVX \footnote{Problem (24) has a solution if for given ${\bf g}_j$, (24) is a feasible problem. The solution from Algorithm 1  ${\bf w}^\star$  can be used as the initial point in the iterative algorithm such that ${\bf X}^{(0)}={\bf w}^\star{\bf w}^{\star H}$. Therefore, a feasible solution of (24) (and also (22)) can always be obtained.}.

The iterative algorithm to solve (22) is summarized in Algorithm 2 (from line 2 to line 7)\footnote{The transmitter may be subject to the transmit power spectrum density constraint. Power spectrum density constraint could refer to the limit of transmit power at each frequency sub-band \cite{Zheng2017}. Power spectrum density constraint can be formulated as $1/2 {\vert s_n\vert}^2 \leq {\bar P}_s\ \forall n$ in problem (13), where ${\bar P}_s$ is the transmit power spectrum density constraint. The constraint can be reformulated as $\mbox{diag}({\bf X})\leq 2{\bar P}_s$ in problem (22) and (24). Problem (24) with the new additional constraint is an SDP problem and can be solved by using CVX. The iterative algorithm to solve (22) with additional power spectrum density constraint is summarized in Algorithm 2.}. Because ${\bf g}_j^{\star}$ is the global optimal solution of problem (23) and problem (24) is convex, it shows that $\gamma^{(l-1)}\geq \gamma^{(l)}$. Note that the stopping criterion of Algorithm 2 is related to the convergence of $\gamma$ (i.e., objective function of problem (22)). As the value of $\gamma$ is monotonically decreasing and bounded, the iterative algorithm is guaranteed to converge. In the Appendix, we show that under the condition that the eigenvector of ${\bf D}$ can be uniquely obtained, the minimizers in Algorithm 2 converges to a KKT point of problem (22).

\begin{algorithm}[t]
\caption{Iterative Algorithm}
\label{Algthm_A2}
\begin{algorithmic}[1]
\State \textbf{Initialize}: $l^{(0)}\gets 0$, ${\bf X}^{(0)}$, $\epsilon$, $\gamma^{(0)}$ and update ${\{{\bf t}_j^{(0)}\}}_{j=1}^K$ by (24d).
\Repeat
	\State $l=l+1$
    \State Find ${\bf g}_j^{\star}$ in (23); ${\bf g}_j^{(l)} \gets {\bf g}_j^{\star}$.
    \State Solve SDP problem in (24); ${\bf X}^{(l)} \gets {\bf X}^\star$, ${\gamma}^{(l)} \gets {\gamma}^\star$.
    \State  Update $t_{j,k}^{(l)}\ \forall\ j,k$ in (24d) and (24e).
	\Until{${\vert{\gamma}^{(l)}-{\gamma}^{(l-1)}\vert}/{\vert{\gamma}^{(l)}\vert}\leq\epsilon$ }.
    \State Find rank-one solution of ${\bf X}^\star$.
\end{algorithmic}
\end{algorithm}
%
Solving the SDP problem (24) by using the interior point method usually yields a high-rank $X^{\star}$ solution \cite{Dattoro}. In the following, in order to obtain the rank-constrained solution that satisfy (20g), we derive a rank-one solution from the high rank ${\bf X}^{\star}$ (line 8 of Algorithm 2).

\subsubsection{Obtaining a Rank-One $\mathbf{X}$}
We first show that if $K\leq 2$ (i.e., $j=1, 2$), (21) can offer a rank-one optimal solution of ${\bf X}^{\star}$ although the solution in (24) has a high rank of ${\bf X}^{\star}$. From \cite{YH}, (21) can be converted into an equivalent form as
\begin {equation}
\begin{aligned}
&\underset{{\bf X}\succeq 0, {\bf g}_j}{\min} &&\Tr\{{{\bf A}{\bf X}}\}\\
&\mbox{s. t.} &&\frac{\Tr{({\bf H}_j^H{\bf g}_{j}{\bf g}_{j}^{H} {\bf H}_j{\bf X})}}{\sigma^2+\Tr{({\bf \tilde H}_j^H{\bf g}_{j}{\bf g}_{j}^{H} {\bf \tilde H}_j{\bf X})}} \geq {\bar \rho}_j,&&&\forall j\\
& &&\Tr\{{\bf X}\}\leq 2P,\\
& &&{\Vert{\bf g}_j\Vert}=1.&&&\forall j
\end{aligned}
\end {equation}
where ${\bf A}={\bf C}+{{\bf C}}^H$ is a Hermitian and ${\bf C}=\sum_{j=1}^Kc_j(-\frac{\beta_2+3/2\beta_4t_{j,0}^{(l-1)}}{4}{\bf M}_{j,0}-3/4\beta_4\sum_{k=1}^{N-1}{[t_{j,k}^{(l-1)}]}^*{\bf M}_{j,k})$. Given an optimized ${\bf g}_j^{\star}$ and by defining ${\bf B}_j=({\bf H}^H{\bf g}_{j}^{\star}{{\bf g}_{j}^{\star}}^{H} {\bf H}-{\bf \tilde H}^H{\bf g}_{j}^{\star}{{\bf g}_{j}^{\star}}^{H} {\bf \tilde H})$, (25) can be recast as
\begin {equation}
\begin{aligned}
&\underset{{\bf X}\succeq 0}{\min} &&\Tr\{{{\bf A}{\bf X}}\}\\
&\mbox{s. t.} &&\Tr{({\bf B}_j{\bf X})}-\sigma^2{\bar \rho}_j\geq 0,&&&\forall j\\
& &&\Tr\{{\bf X}\}\leq 2P,\\
\end{aligned}
\end {equation}
where ${\bf B}_j$ is Hermitian. According to \cite{Huang2010}, (26) is shown to be a separable SDP. By applying [30, Proposition $3.5$], we can show that (26) can yield a rank one optimal solution when $K\leq 2$. Accordingly, a rank one ${\bf X}_{r1}^*={\bf w}^{\star}{\bf w}^{\star H}$ can be obtained by using Rank Reduction (RR) \cite{Huang2010}. Hence, we propose an algorithm that first iteratively optimizes ${\bf g}_j$ in (23) and ${\bf X}$ in (24). RR precedure is then performed to the high rank solution of ${\bf X}^\star$ to find ${\bf w}^{\star}$.

When $K\geq 3$, (26) may not have a rank-one optimal solution of ${\bf X}^{\star}$. In this scenario, we solve (21) by iteratively optimizing ${\bf g}_j$ in (23) and ${\bf X}$ in (24) until convergence. Then, rank randomization method is performed to obtain $T$ rank-one solution from higher rank of ${\bf X}^{\star}$. By taking the eigenvalue decomposition (EVD) ${\bf X}^{\star}={\bf U}{\bf \Sigma}^{1/2}{\bf U}^H$, we generate $T$ random vectors ${\bf w}_t={\bf U}{\bf \Sigma}^{1/2}{\bf v}_t$ for $t=1,..,T$ where ${\bf v}_t$ is a random vector with each complementary entry is from a circular uniform distribution such that $E{[{{\bf v}_t{\bf v}_t^H}]}={\bf I}$. For each ${\bf w}_t$, we find their corresponding ${\bf g}_{t,j}$ from (15). Consequently, the best feasible rank-one solution satisfying constraint (20c) and (20f) is obtained as follows. Among $T$ ${\bf w}_t$ and their corresponding ${\bf g}_{t,j}$, we find ${\bf w}_t^{'}$ and ${\bf g}_{t,j}^{'}$ that can satisfy constraint (20c) and (20f). Accordingly, the best feasible solution is ${\bf w}^{\star}=\argmax_{{\bf w}_{t}^{'}} Z_{DC}{({\bf w}_{t}^{'})}$ where ${\bf X}_{r1}^{\star}={\bf w}^{\star}{\bf w}^{\star H}$ and ${\bf g}_j^{\star}$ is given as ${\bf g}_{t,j}^{'}$ that associate with ${\bf w}^{\star}$.

\subsubsection{A Simplified Algorithm} A simplified iterative algorithm with a lower computational complexity than Algorithm 2 is proposed. Given an initial value of ${\bf X}$ with its corresponding ${\bf g}_{j}^{*}$ from (23), by fixing ${\bf g}_{j}^{*}$ at each iteration, (24) is optimized iteratively until convergence. Then, a rank-one ${\bf X}$ is obtained as discussed in section III-B-1. The iterative algorithm is shown in Algorithm 3. For given ${\bf g}_j^{\star}$ in line 2, the iterative algorithm converges to a stationary point of (24) \cite{YH}. Although this algorithm provides a lower complexity compared to Algorithm 2, the performance of Algorithm 2 is expected to be better. This is because alternating optimization that optimizes ${\bf g}_j$ (in line 4 of Algorithm 2) and ${\bf X}$ iteratively in Algorithm 2 can reduce the value of $\gamma$ over the iterations.

\begin{algorithm}[t]
\caption{A Simplified Algorithm}
\label{Algthm_A3}
\begin{algorithmic}[1]
\State \textbf{Initialize}: $l^{(0)}\gets 0$, ${\bf X}^{(0)}$, $\epsilon$, $\gamma^{(0)}$ and update ${\{{\bf t}_j^{(0)}\}}_{j=1}^K$ by (24d).
\State Find ${\bf g}_{j}^{\star}$ in (23) for given ${\bf X}^{(0)}$.
\Repeat
	\State $l=l+1$
    \State Solve SDP problem in (24); ${\bf X}^{(l)} \gets {\bf X}^\star$, ${\gamma}^{(l)} \gets {\gamma}^\star$.
    \State  Update $t_{j,k}^{(l)}\ \forall\ j,k$ by (24d) and (24e).
\Until{${\vert{\gamma}^{(l)}-{\gamma}^{(l-1)}\vert}/{\vert{\gamma}^{(l)}\vert}\leq\epsilon$ }.
\State Find rank-one solution of ${\bf X}^\star$.
\end{algorithmic}
\end{algorithm}
%

\section{Simulation Results}
We consider a system with a centre frequency of 5.18 GHz, 10 MHz bandwidth, 36 dBm EIRP, 2 dBi receive and transmit antenna gain at the tag and 2 dBi receive antenna gain at the reader. The path loss for forward and backward link are 58 dB and the NLOS channel power delay profile is obtained from model B \cite{Medbo1998}. The channel taps each with an average power $\beta_l$, are independent, circularly symmetric complex random Gaussian distributed and normalized such that $\sum_l \beta_l=1$. This leads to an average receive power of -20 dBm at the tag. The SNR at the reader is defined as $P/\sigma^2$. In the simulation, it is assumed that each tag has the same SINR constraint ${\bar \rho}$. For simplicity the weight for energy harvested $c_j=1$ for all tags.

\begin{figure}[!t]
\centering
\includegraphics[width=4in]{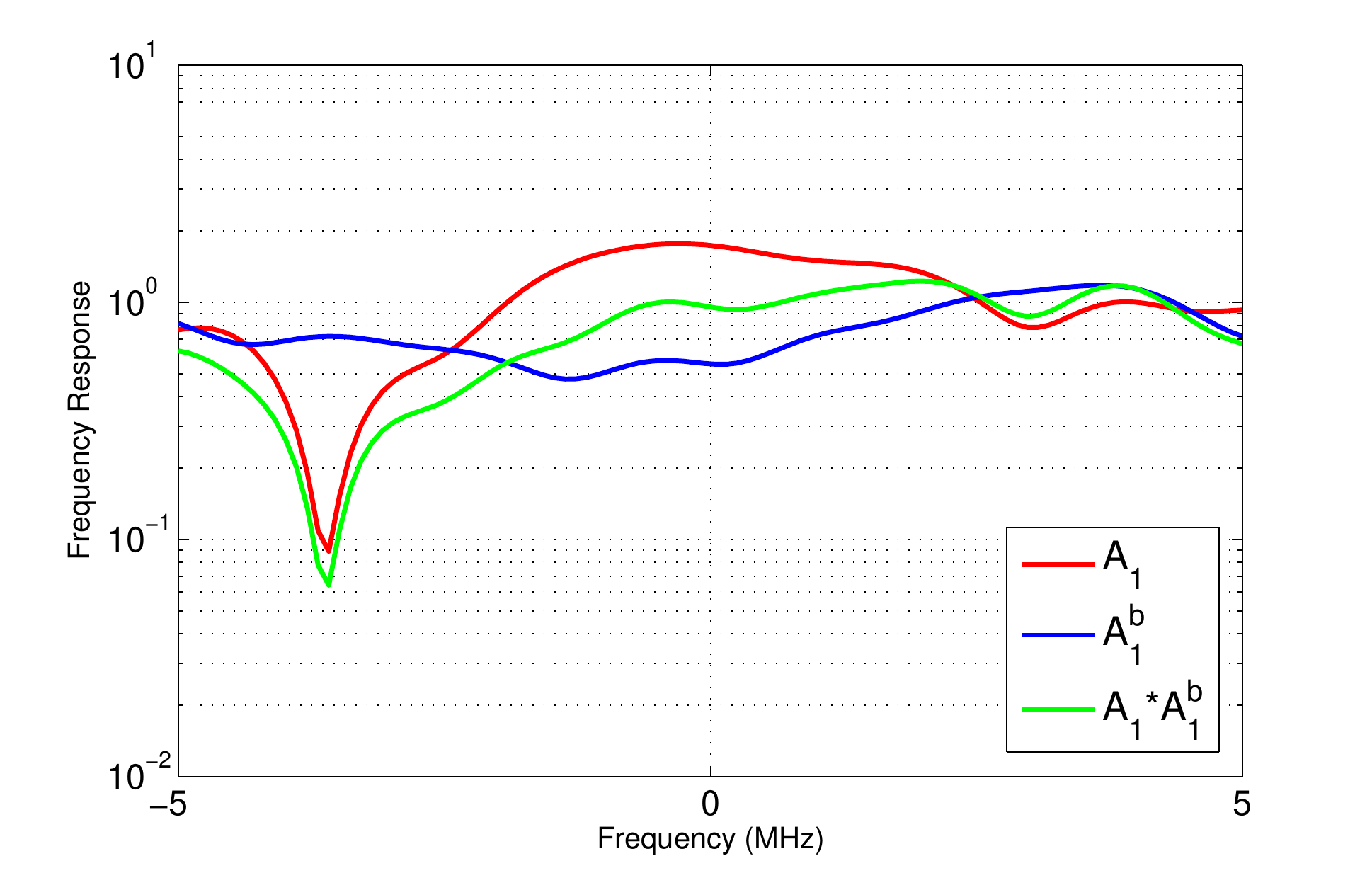}
\caption{Frequency response of the channel for $K=1$.}
\label{channelK1}
\end{figure}

\begin{figure}[tbh]
    \subfigure[SNR=5dB.]
            {\includegraphics[width=3.7in]{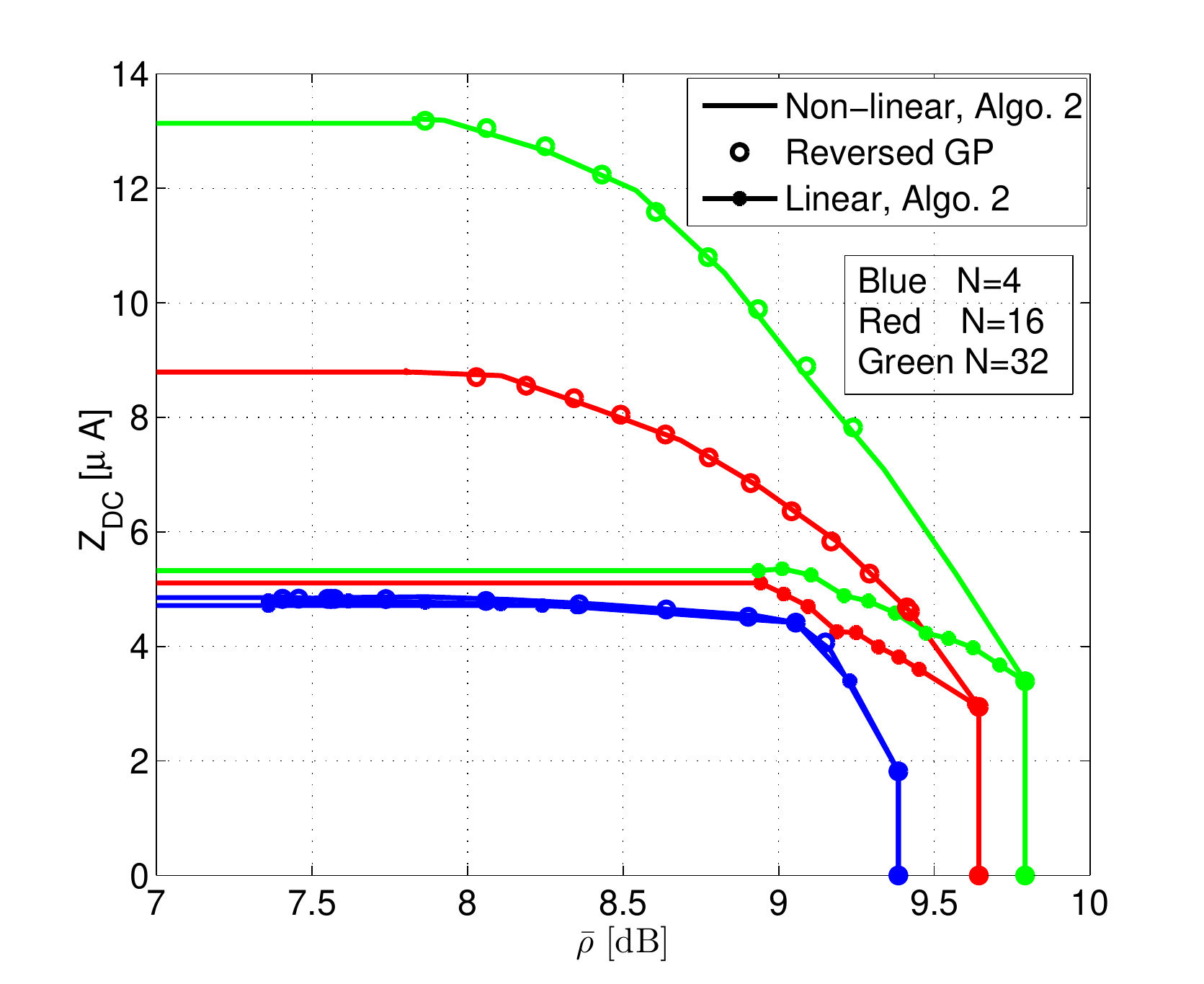}}
    ~
    \subfigure[SNR=20dB.]
        {\includegraphics[width=3.7in]{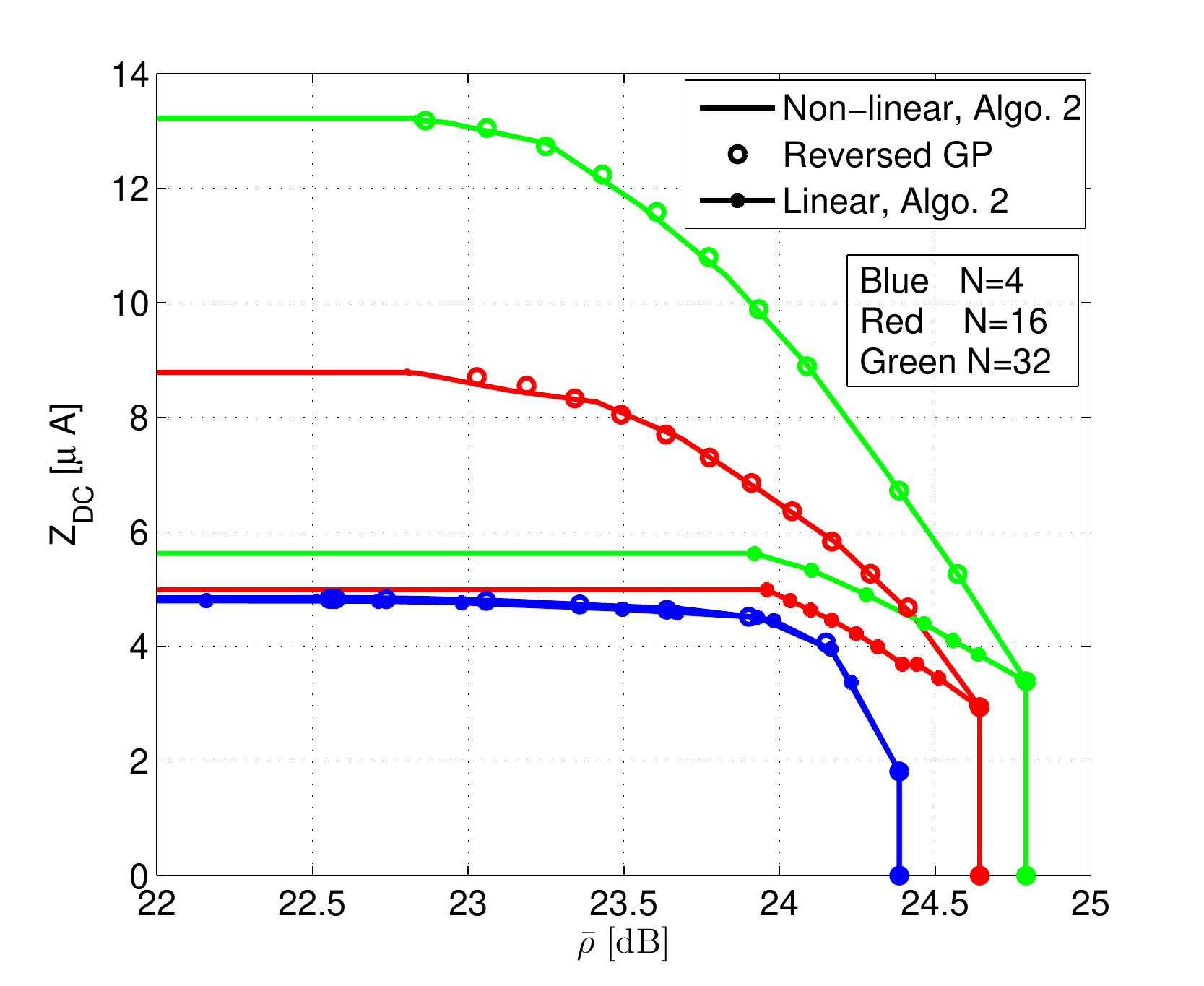}}
    \caption{$\bar \rho$ -$Z_{DC}$ region for $K=1$.}
    \label{tradeK1}
\end{figure}

\begin{figure}[tbh]
    \subfigure[Frequency response of forward channel.]
        {\includegraphics[width=3.7in]{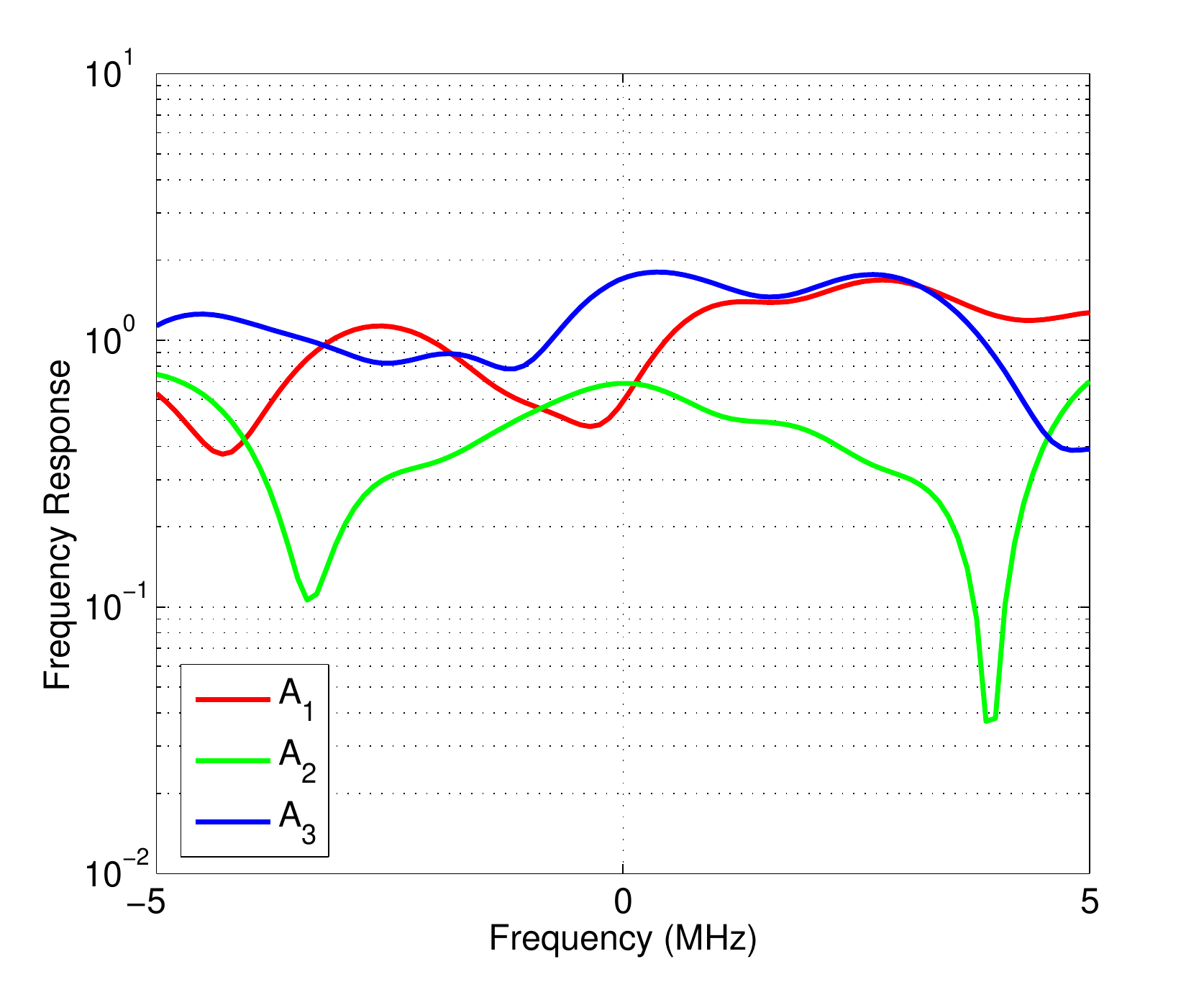}}
    ~
    \subfigure[Frequency response of backscatter channel.]
        {\includegraphics[width=3.7in]{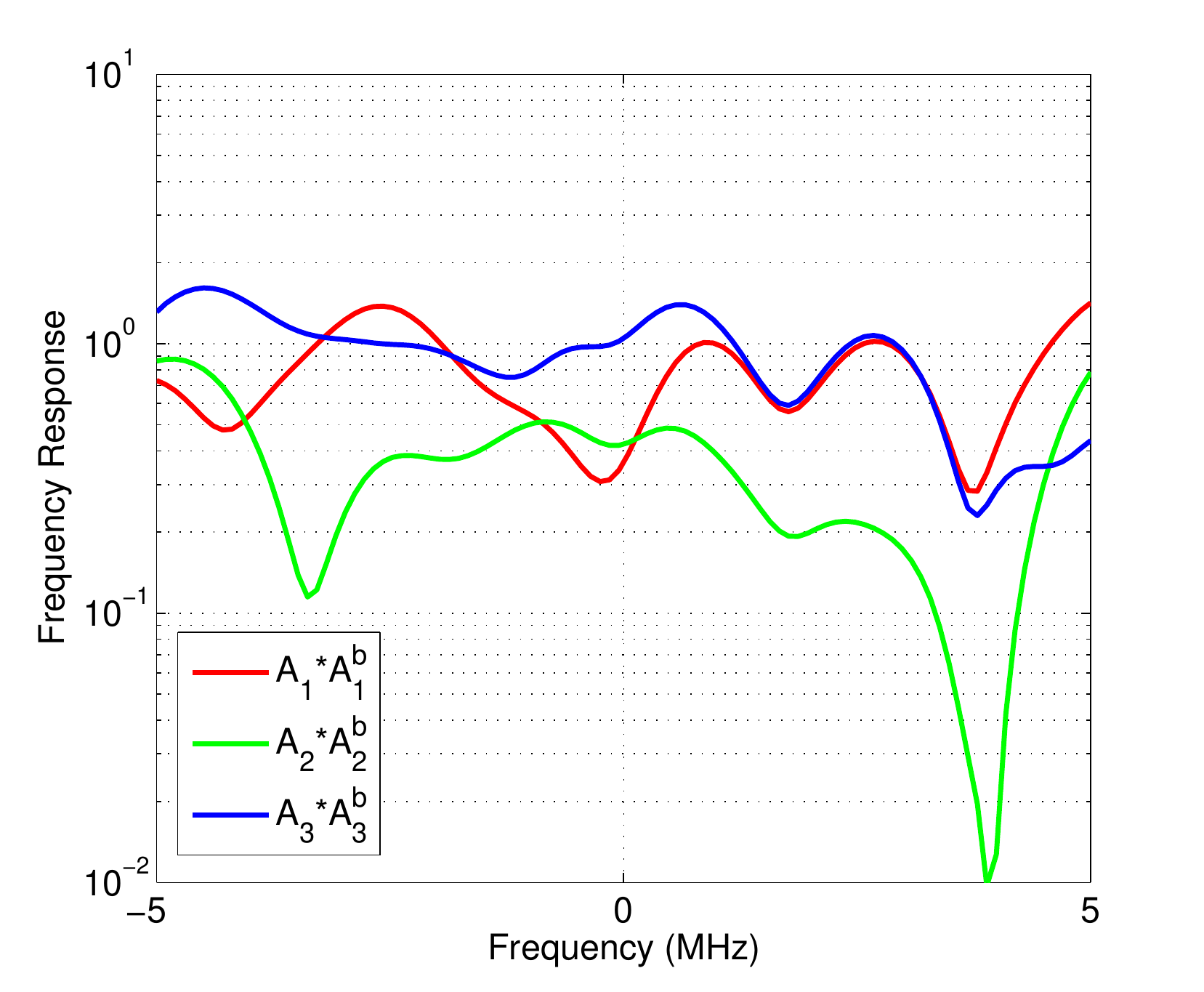}}
    \caption{Frequency response of the channel for $K=3$.}
    \label{channelK3}
\end{figure}

\begin{figure}[tbh]
    \subfigure[SNR=5dB.]
        {\includegraphics[width=3.7in]{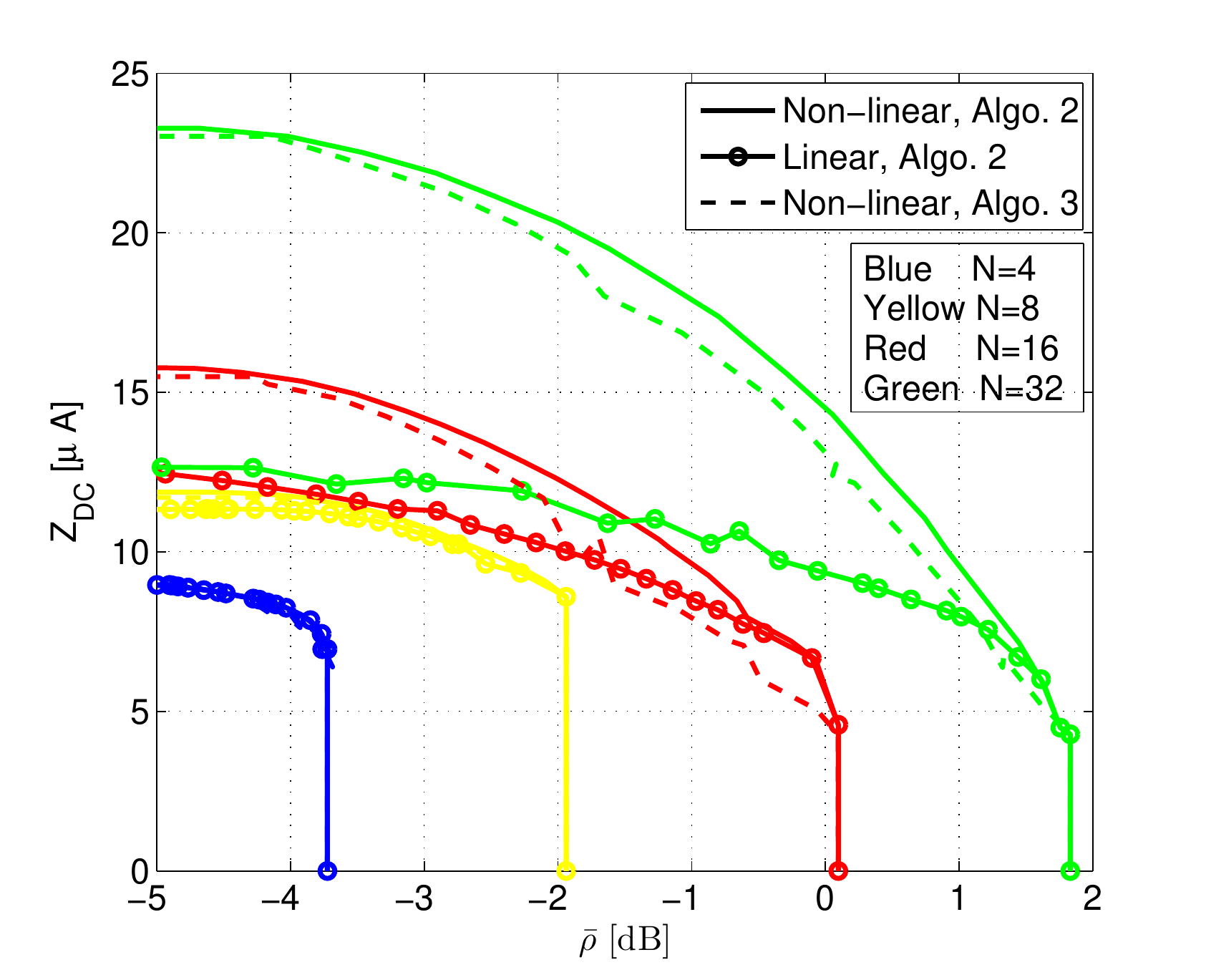}}
    ~
     \subfigure[SNR=20dB.]
        {\includegraphics[width=3.7in]{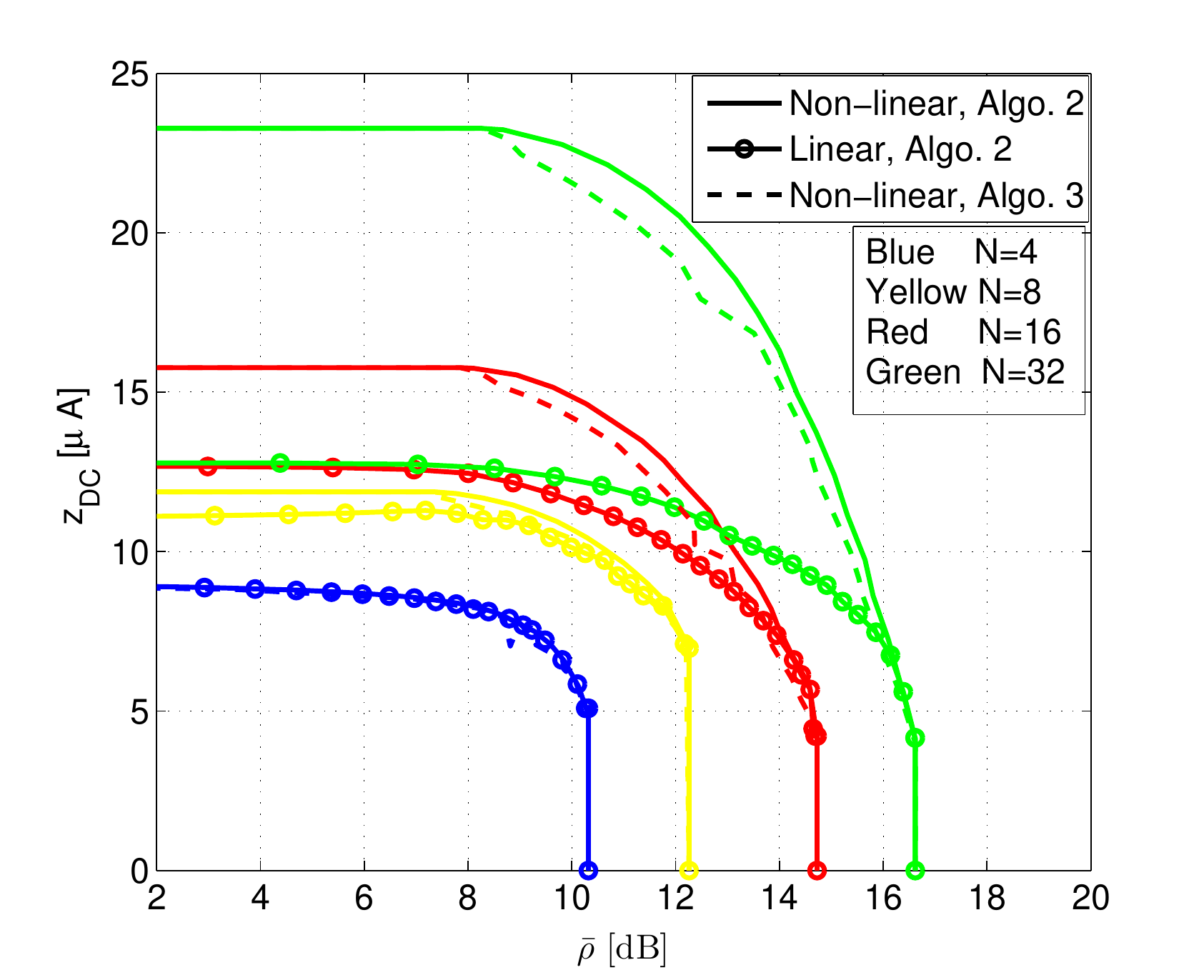}}
    \caption{$\bar \rho$ -$Z_{DC}$ region for $K=3$.}
    \label{tradeK3}
\end{figure}

Simulation for $K=1$ and $K=3$ are run over a channel realization for two different SNRs (5 and 20 dB). The channel frequency responses are illustrated in Fig. \ref{channelK1} for $K=1$  and Fig. \ref{channelK3} for $K=3$. The ${\bar \rho}$-$Z_{DC}$ region for $K=1$ and $K=3$ are shown in Fig. \ref{tradeK1} and Fig. \ref{tradeK3}, respectively. The $Z_{DC}$ on the y-axis is given in (17). The SINR (SNR for a point-to-point system) on the x-axis is the SINR constraint ${\bar \rho}$. The blue, red and green colours in the figures refer to $N=4, 16$ and $32$, respectively. The solid lines in the figures indicate the result for waveform optimization based on nonlinear EH model and obtained by using Algo. 2. In point-to-point system where $K=1$, there is no interference signal. The maximum achievable SNR in Fig. \ref{tradeK1} are obtained by allocating full transmit power to a single subcarrier corresponding to $\argmax_n{\vert{h}_{1,n}{h}_{1,n}^b\vert}$. On the other hand, the maximum $Z_{DC}$ can be obtained from \cite{Clerckx2016, YH}. For $K=3$, the extreme point on the x-axis and y-axis in Fig. \ref{tradeK3} are obtained from Algo. 1 and \cite{ YH}, respectively. 

Note that for a point-to-point system, the phase of the transmit waveform can be obtained in a closed-form as in \cite{Clerckx2017} \footnote{However, closed-form solution cannot be obtained when $K>1$.}. Accordingly, power allocation across frequency component can be optimized by using Reversed GP \cite{Clerckx2017}. It is observed that the achievable $\bar \rho$-$Z_{DC}$ regions obtained by using Reversed GP and Algo. 2 are comparable as shown in Fig. \ref{tradeK1} \footnote{Due to the non-convexity of the problem, the solutions from Reversed GP and Algorithm 2 cannot guarantee to converge to the global optimal solution. However, both approaches converge to a KKT point.}. Table \ref{table1} compares the average computational complexity for waveform optimization based on Algo. 2 and Reversed GP. $Z_{DC}$ and the elapsed time are averaged over several channel realizations with $N=4, 8$. To draw the comparison between the Reversed GP approach and Algo. 2, we employ the same initial point and $\epsilon=10^{-7}$ for both algorithms and the SNR constraint at the tag $\bar \rho= 10 dB$. The simulation is conducted by MATLAB R2013b on a computer with an Intel Core i7 processor at 3.3GHz, RAM of 16GB and Windows 10 Pro. The simulation results show that the average elapsed time for Algo. 2 is significantly less than Reversed GP, while the $Z_{DC}$ value from both algorithms is comparable. In terms of asymptotic computational complexity, the Reversed GP algorithm suffers from exponential complexity \cite{Chiang} to compute an optimal solution while solving SDP by using interior point method may take polynomial complexity \cite{SDPComplexity} per iteration in Algorithm 2.

\begin{table}[]
\centering
\caption{Elapsed running time:Algorithm 2 vs. Reversed GP}
\label{table1}
\includegraphics[width=3.7in]{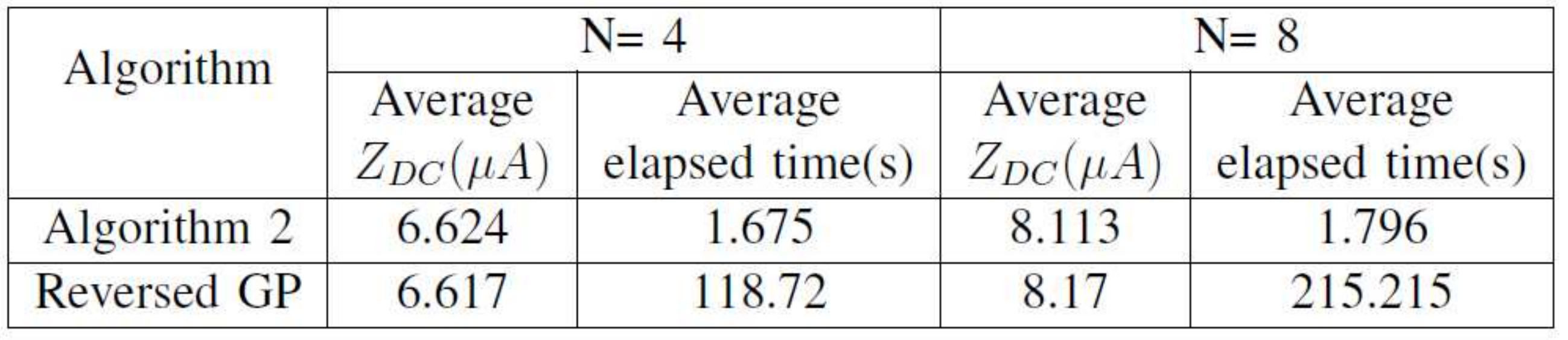}
\end{table}

From Fig. \ref{tradeK1} and Fig. \ref{tradeK3}, by looking at the ${\bar \rho}$-$Z_{DC}$ region for the nonlinear EH model obtained from Algo. 2, it is observed that the performance of wirelessly powered backscatter communication is subject to a tradeoff between $Z_{DC}$ and $\bar \rho$. That is, maximizing the $Z_{DC}$ would result in a decrease in $\bar \rho$. The second observation is that by looking at two different SNRs ($5$ and $20$ dB), the ${\bar \rho}$-$Z_{DC}$ tradeoff slope (i.e., rate of change) decreases as the value of SNR increases, for each $N$. Another observation is that, the tradeoff region can be enlarged as the number of sinewaves in the waveform increases. This is because a proper power allocation across multiple sinewaves can exploit a frequency diversity gain and the nonlinearity of the rectifier \cite{Clerckx2017, Clerckx2016}.

We then study the performance of the optimized waveform based on the linear EH model which accounts only for the second order term in the Taylor expansion (i.e., the second order term in (6)). By taking only the second order term in (17), $Z_{DC}=\sum_{j=1}^K c_j\beta_2t_{j,0}$ and constraint (24b) become $-\sum_{j=1}^K c_j\beta_2t_{j,0}-\gamma \leq 0$. Accordingly, we can find that (24) is a SDP problem. Therefore, waveform optimization based on the linear EH model in (13) can be solved by using Algo. 2. It can be drawn from Fig. \ref{tradeK1} and Fig. \ref{tradeK3} that for small $N$ (e.g., $N=4$), the linear model-based design benefits from channel frequency selectivity to get close performance to the nonlinear model-based design. As $N$ increases, the waveform design based on the nonlinear model explicitly outperforms the linear model. The performance gap of the linear model-based design compared to the nonlinear model-based design increases as $N$ increases. It is observed that as $N$ keeps increasing ($N\geq 16$), the tradeoff region for the linear model does not change much. This is because the linear model provides only a logarithmic increase of $Z_{DC}$ with $N$ \cite{Clerckx2016}. These observations show the importance of accounting for the nonlinearity of the rectifier to increase the efficiency of the backscatter communication waveform design.

The performance of the waveform optimized based on the simplified algorithm as described in Algo. 3 for $K=3$ is illustrated in Fig. \ref{tradeK3}. It is observed that $\bar{\rho}$-$Z_{DC}$ region offered by Algo. 3 is slightly smaller than Algo. 2, with a lower computational complexity. Table \ref{table2} compares the average computational complexity for waveform optimization based on Algo. 2 and Algo. 3. $Z_{DC}$ and the elapsed time are averaged over several channel realizations with $(K,N)$ is set to $(4,8)$ and $(6,64)$. To draw the comparison, we employ the same initial point and $\epsilon=10^{-7}$ for both algorithms and the SINR constraint at each tag $\bar \rho$ is  set to $3 dB$. The simulation is conducted by using the same MATLAB version and computer as in Table \ref{table1}. The simulation results show that the average elapsed time for Algo. 3 is less than Algo. 2, with the $Z_{DC}$ achieved from Algo. 3 is slightly lower than Algo. 2.

\begin{table}[]
\centering
\caption{Elapsed running time:Algorithm 2 vs. Algorithm 3}
\label{table2}
\includegraphics[width=3.7in]{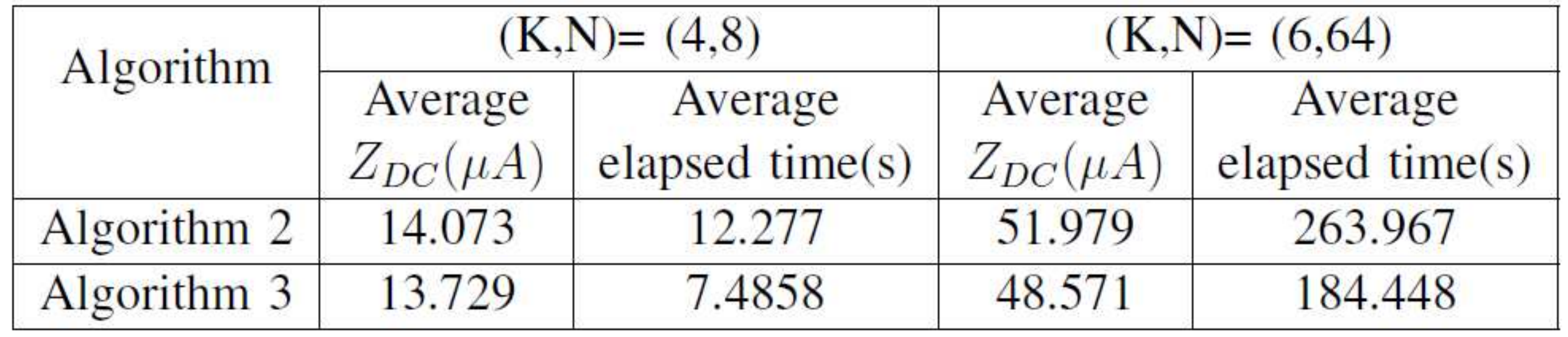}
\end{table}

\begin{figure}[!t]
\centering
\hspace{-0.3cm}
\includegraphics[width=4in]{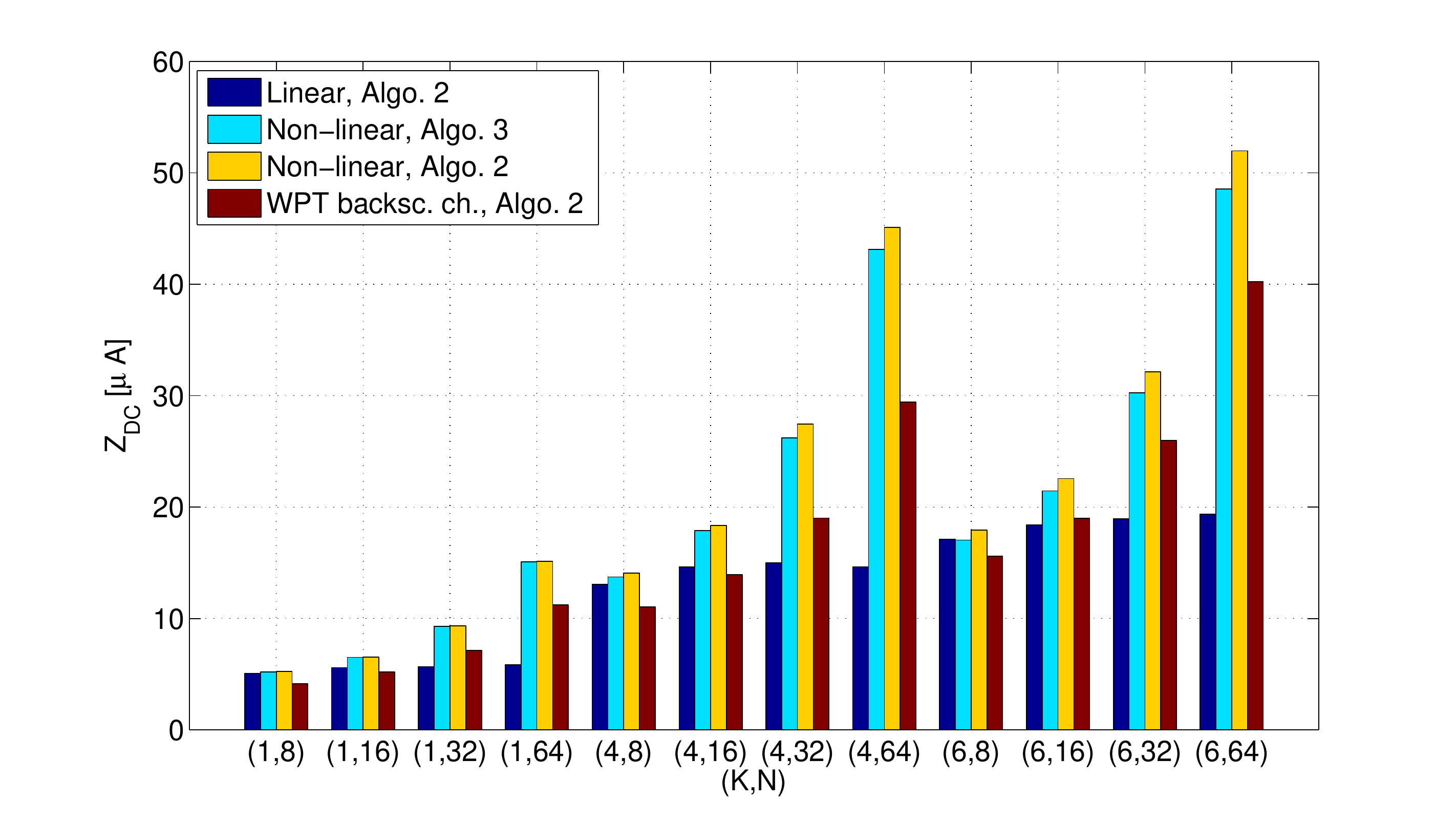}
\caption{$Z_{DC}$ with ($K,N$) and $\bar \rho=3dB$.}
\label{Fig8}
\end{figure}

Fig. \ref{Fig8} shows the performance of average $Z_{DC}$ under different $N$ and $K$. Simulation for both schemes are run over several channel realization SNR 20dB. SINR constraint at the reader for each tag is fixed to 3dB. The observations from Fig. \ref{Fig8} are as follows. {\it First}, waveform design based on the nonlinear model has been found to outperform that based on the linear model for all configurations. {\it Second}, the waveform optimized by using Algo. 3 has close performance to those obtained by using Algo. 2. {\it Third}, waveform design benefits from a multiuser diversity gain to exhibit a significant increment of $Z_{DC}$ with the number of tags. Nevertheless, the average amount of energy harvested at each tag decreases as the number of tag increases. This is because, the waveform has to be designed to meet all SINR requirements, and therefore decreases the amount of energy harvested  at each tag. It is also observed that for a given $N$, the performance gap between waveform optimized based on the nonlinear model and the linear model increases as $K$ increases. {\it Fourth}, waveform design benefits from a combined frequency diversity gain and the rectifier nonlinearity gain to exhibit a larger $Z_{DC}$ with $N$. It can be observed that the inefficiency of the linear model is more severe as $N$ increases irrespective of $K$. Fifth, we investigate the performance of $Z_{DC}$ when the waveform is optimized by using backscatter channel to maximize the energy harvested \footnote{Since CSIT of forward channel is difficult to attain in a backscatter system, the power transfer optimization based on the backscatter channel has been considered as in \cite{Gyang2015, Arnitz2013}.}. Here, (20) is optimized by substituting forward channel $h_{j,n}$ with backscatter channel $h_{j,n}h_{j,n}^b$ in (17). It is observed that waveform optimized by using backscatter channel to maximize the energy harvested leads to a smaller $Z_{DC}$ than those optimized by using forward channel, the difference is severe as $N$ increases.

\begin{figure}[tbh]
    \subfigure[Average $Z_{DC}$.]
        {\includegraphics[width=4in]{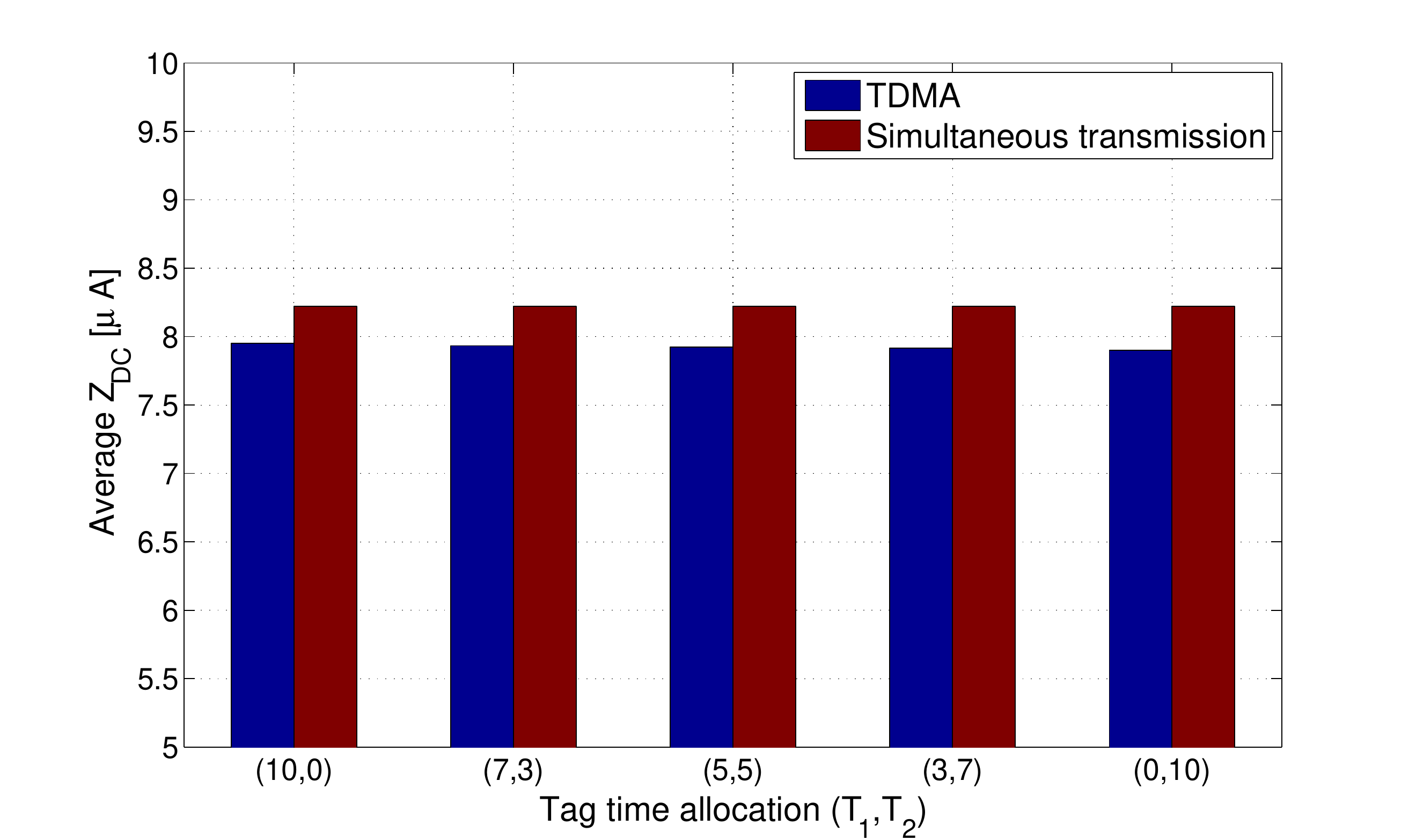}}
        \label{Fig9a}
    ~
     \subfigure[Average $z_{DC}$ for each user.]
        {\includegraphics[width=4in]{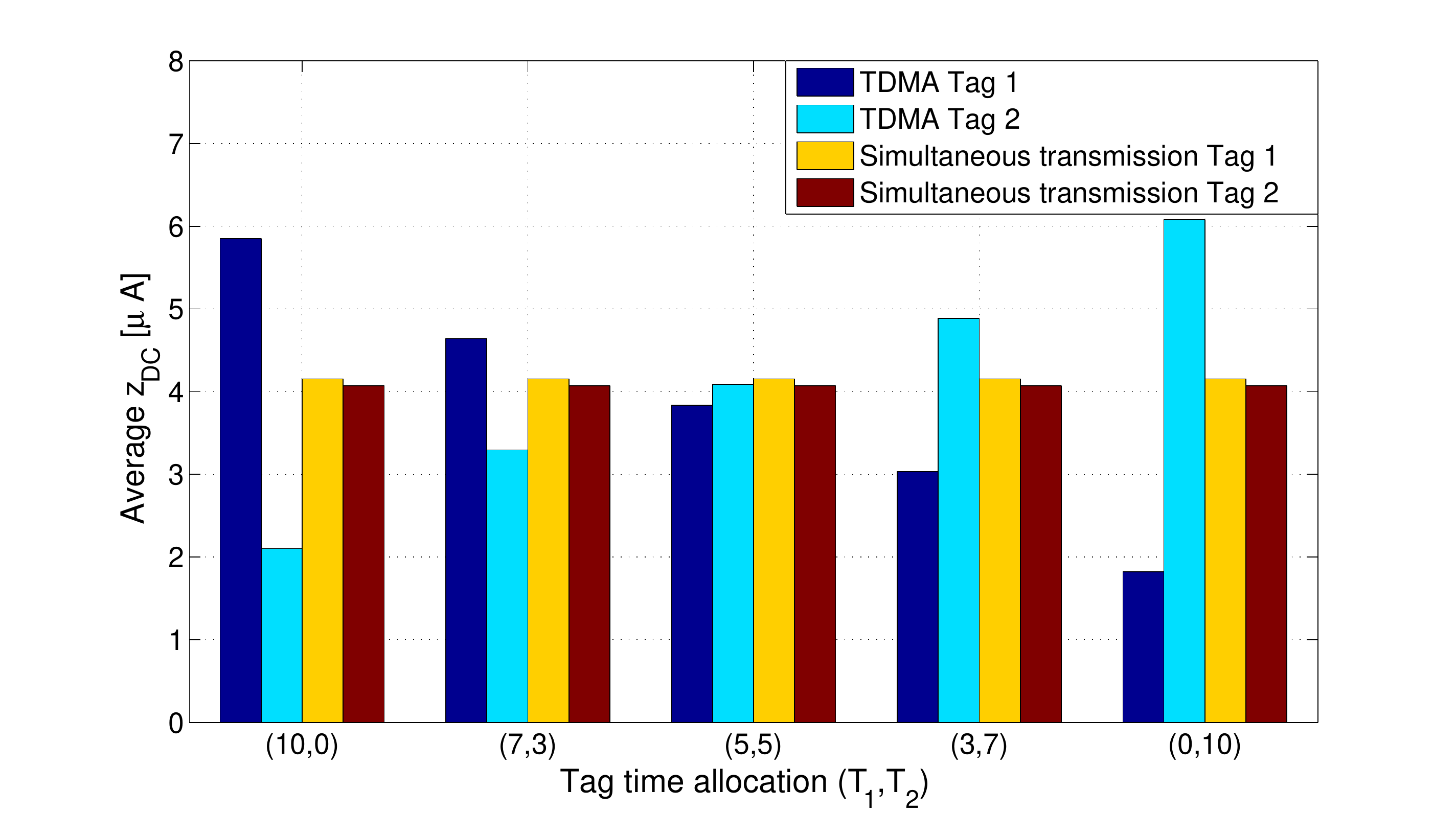}}
        \label{Fig9b}
    \caption{Comparison of the simultaneous transmission and the TDMA scheme.}
    \label{Fig9}
\end{figure}

We then compare the average $Z_{DC}$ performance achieved by the simultaneous transmission scheme which is formulated as (13) and the TDMA scheme. Simulation of both schemes are run for $K=2$, $N=8$ and SNR $P/{\sigma}^2=10$dB. The SINR constraint ${\bar \rho}_j$ is set to $10$dB at each tag for both the simultaneous transmission and the TDMA scheme. In the simultaneous transmission, two tags (i.e. tag 1 and tag 2) are simultaneously served by the RF transmitter/reader, harvesting wireless power and delivering backscattered signals. By contrast, in the TDMA scheme, tag 1 and tag 2 are separately served in various time slots. For the TDMA scheme, let ($T_1$,$T_2$) represents the number of time slots allocated to tag 1 and tag 2, respectively. In the simulation, we consider that $T_1+T_2=10$. As depicted in Fig. \ref{Fig9a}, the average $Z_{DC}$ ($Z_{DC}=\sum_{j=1}^Kz_{DC,j}$, the sum of energy harvested at both tag) performance is obtained by averaging $Z_{DC}$ over random channel realizations. Note that in Fig. \ref{Fig9a}, the notations ($T_1$,$T_2$) are associated with only the (blue) bars for the TDMA schemes, while the (red) bar for the simultaneous transmission is repeated for a visual comparison. It can be drawn from Fig. \ref{Fig9a} that the simultaneous transmission can outperform the TDMA scheme, yielding a higher average $Z_{DC}$. It is noteworthy that in contrast to the TDMA scheme, the simultaneous transmission enables all tags to transmit data at the same time. Fig. \ref{Fig9b} demonstrates the average $z_{DC}$ (energy harvested at the tag) achieved by each tag. It is shown that compared to the TDMA schemes where $T_1 \neq T_2$, the simultaneous transmission can provide a fairness-aware solution, in terms of the average $z_{DC}$ achieved by each tag. We also make the observation that for TDMA scheme where $T_1 = T_2$, the average $z_{DC}$ achieved by each tag is lower than that achieved by the simultaneous transmission\footnote{Note that if the number of random channel realizations is adequately large and all the tags suffer the same large-scale fading, the tags should achieve the same average $z_{DC}$ for the simultaneous transmission (or a TDMA scheme where $T_1 = T_2$).}. Such an observation confirms that the average $z_{DC}$ achieved by each tag can benefit from the simultaneous transmission.

\section{Conclusion}
This paper studies waveform design for wirelessly powered backscatter communication in a multiuser backscatter system. The waveform is designed to be adaptive to the CSI of forward and backward channels available at the RF transmitter. The tradeoff between harvested energy at the tags and SINR of the backscatter communication is investigated. An efficient algorithm is derived to optimize the transmit waveform and receive combiner so as to identify the tradeoff region. The performance of optimized waveforms based on linear and nonlinear EH models are studied. Numerical results show waveform design based on the nonlinear EH model which accounts for the nonlinearity of the rectifier provides a significant gain over those obtained based on the linear EH model. It was observed that power allocation across multisine waveform and adaptive to the CSIT are beneficial to enlarge the tradeoff region. Accounting for multiuser backscatter, the system benefits from the multiuser diversity to increase the total amount of energy harvested in the system for a given SINR constraint. Nevertheless, the amount of energy harvested at each tag decreases as the waveform has to be designed to meet the SINR constraints of all tags.

\appendix
\section{Appendix A}
In this section, we show that under the condition that the eigenvector of ${\bf D}$ can be uniquely obtained, the minimizers in Algorithm 2 converges to a KKT point of problem (22).
Let represent $\{\gamma, {\bf X}, {\bf t}_j\}=\Psi({\bf g}_j)$ and ${\bf g}_j=\Theta({\bf X})$ as optimization problem (23) and (24), respectively. It is noted that the value of $\gamma$ is monotonically decreasing due to optimality of $\Psi(.)$ and $\Theta(.)$. Theoretically, the eigenvector of ${\bf D}$ is not unique such that the minimizers ${\bf X}$ and ${\bf g}_j$ may not converge. However, similar to \cite{YH} and \cite{Bezdek2003}, under the assumption that the eigenvector of ${\bf D}$ can be uniquely obtained, the sequence of minimizer ${\{\gamma^{(l)}, {\bf t}_j^{(l)}, {\bf X}^{(l)}, {\bf g}_j^{(l)}\}}_{l=0}^\infty$ converges to a limit point ${\{{\hat \gamma}, {\bf {\hat t}}_j, {\bf {\hat X}}, {\bf {\hat g}}_j\}}$. Alternatively, the stopping criterion of Algorithm 2 can also be related to the convergence of the minimizers such as ${\Vert {\bf X}^{(l)}-{\bf X}^{(l-1)}\Vert}_F/ {\Vert {\bf X}^{(l)}\Vert}_F \leq \epsilon$.

Note that problem (22) is similar to problem (24) except that ${\bf g}_j$ which is found in constraint (20c) and (20h) is optimized in problem (23). For brevity, here we discuss the KKT conditions associated with constraint (20c) and (20h). The remaining KKT conditions of (22) is not discuss since it can be easily found that the the remaining KKT conditions for problem (22) are also the KKT conditions for problem (24).

Let $\{{ \gamma}^{}, {\bf { t}}_j^{}, {\bf { X}}^{}, {\bf { g}}_j^{}\}$ and Lagrangian multipliers $\{{ \mu}^{}, { { \nu}}_j^{}, { { \xi}}_{j,k}^{}, {\eta}_{j,k}^{}, { { \varpi}}^{}, {\kappa}_j^{}\}$ associated with constraints $\{(22b), (20c), (20d), (20e), (20f), (20h)\}$ satisfy the KKT conditions of problem (22). 
Let $f_{ j,k}^{'}({\bf X}, t_{j,k})=\Tr\{{\bf M}_{j,k}{\bf X}\}-t_{j,k}$, $f_{j,k}^{''}({\bf X}, t_{j,k})=\Tr\{{\bf M}_{j,k}^H{\bf X}\}-t_{j,k}^*$ and $f^{'''}({\bf X})=\Tr\{{\bf X}\}-2P$. From the Lagrangian of problem (22), the KKT conditions associated with constraint (20c) and (20h) are as follows
\begin{subequations}
\begin{align}
&\sum_{j=1}^K{ { \nu}}_j^{} \nabla_{{\bf X}}\rho_j({\bf X}^{}, {\bf g}_j^{})+\sum_{j=1}^K\sum_{k=0}^{N-1}{ { \xi}}_{j,k}^{}\nabla_{{\bf X}}f_{j,k}^{'}({\bf X}^{}, t_{j,k}^{})+\nonumber\\ &\sum_{j=1}^K\sum_{k=1}^{N-1}{\eta}_{j,k}^{}\nabla_{{\bf X}}f_{j,k}^{''}({\bf X}^{}, t_{j,k}^{})+{ { \varpi}}^{}\nabla_{{\bf X}}f^{'''}({\bf X}^{})=0,\\
&\nu_j^{} \nabla_{{\bf g}_j}\rho_j({\bf X}^{}, {\bf g}_j^{})=0,\ \ \ \ \ \ \ \ \ \ \ \ \ \ \ \ \ \ \ \ \ \ \ \ \ \ \ \ \ \ \ \ \ \forall j\\
&0\leq { { \nu}}_j^{} \perp \rho_j({\bf X}^{}, {\bf g}_j^{})-{\bar \rho}_j\geq 0\ \ \ \ \ \ \ \ \ \ \ \ \ \ \ \ \ \ \ \ \ \ \ \ \forall j\\
&\Vert {\bf g}_j^{}\Vert = 1\ \ \ \ \ \ \ \ \ \ \ \ \ \ \ \ \ \ \ \ \ \ \ \ \ \ \ \ \ \ \ \ \ \ \ \ \ \ \ \ \ \ \ \ \ \ \ \ \forall j
\end{align}
\end{subequations}

Based on problem (24),  $\{{\hat \gamma}, {\bf {\hat t}}_j, {\bf {\hat X}}\}$ with Lagrangian multipliers $\{({ {\hat \mu}}, { { {\hat \nu}}}_j, { { {\hat \xi}}}_{j,k}, {{\hat \eta}}_{j,m}, { { {\hat \varpi}}})\}$ associated with constraints $\{(24b), (24c), (24d), (24e), (24f)\}$ must satisfy the KKT conditions of problem (24). The KKT conditions of problem (24) associated with constraint (20c) and (20h) are 
\begin{subequations}
\begin{align}
&\sum_{j=1}^K{ {\hat \nu}}_j \nabla_{{\bf X}}\rho_j({\bf \hat X}, {\bf \hat g}_j)+\sum_{j=1}^K\sum_{k=0}^{N-1}{ {\hat \xi}}_{j,k}\nabla_{{\bf X}}f_{j,k}^{'}({\bf \hat X}, {\hat t}_{j,k})+\nonumber\\ &\sum_{j=1}^K\sum_{k=1}^{N-1}{\hat \eta}_{j,k}\nabla_{{\bf X}}f_{j,k}^{''}({\bf \hat X}, {\hat t}_{j,k})+{ {\hat \varpi}}\nabla_{{\bf X}}f_{j,k}^{'''}({\bf \hat X})=0,\\
&0\leq { { {\hat \nu}}}_j \perp \rho_j({\bf {\hat X}}, {\bf {\hat g}}_j)-{\bar \rho}_j\geq 0.\ \ \ \ \ \ \ \ \ \ \ \ \ \ \ \forall j
\end{align}
\end{subequations}

Based on problem (23), ${\bf {\hat g}}_j$ must satisfy the KKT conditions of problem (23). The KKT conditions of problem (23) are
\begin{subequations}
\begin{align}
&\nabla_{{\bf g}_j}\rho_j({\bf {\hat X}}, {\bf {\hat g}}_j)=0,&&\forall j\\
&\Vert {\bf \hat g}_j\Vert = 1.&&\forall j
\end{align}
\end{subequations}
Multipying (29a) by ${\hat \nu}_j$, the KKT conditions in (29) are given as
\begin{subequations}
\begin{align}
&{\hat \nu}_j \nabla_{{\bf g}_j}\rho_j({\bf {\hat X}}, {\bf {\hat g}}_j)=0,&&&\forall j\\
&\Vert {\bf \hat g}_j\Vert = 1.&&&\forall j
\end{align}
\end{subequations}
By combining the KKT conditions of (28) and (30), we can obtain the KKT conditions as in (27). Hence, ${\{{\hat \gamma}, {\bf {\hat t}}_j, {\bf {\hat X}}, {\bf {\hat g}}_j\}}$ is the KKT point of problem (22).

\begin{IEEEbiography}[{\includegraphics[width=1in,height=1.25in,clip,keepaspectratio]{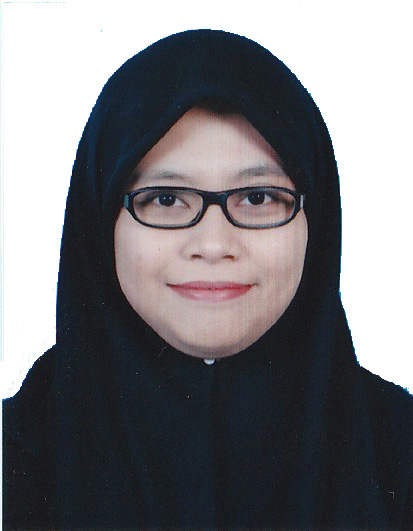}}]{Zati Bayani Zawawi}
Zati Bayani Zawawi received the M.Sc degree in mobile communication systems from University of Surrey, UK in 2012 and the Ph.D. degree in electrical engineering from Imperial College London, U.K. in 2018. Her research interests include wireless communications, wireless power transfer and wireless powered communications.
\end{IEEEbiography}

\begin{IEEEbiography}[{\includegraphics[width=1in,height=1.25in,clip,keepaspectratio]{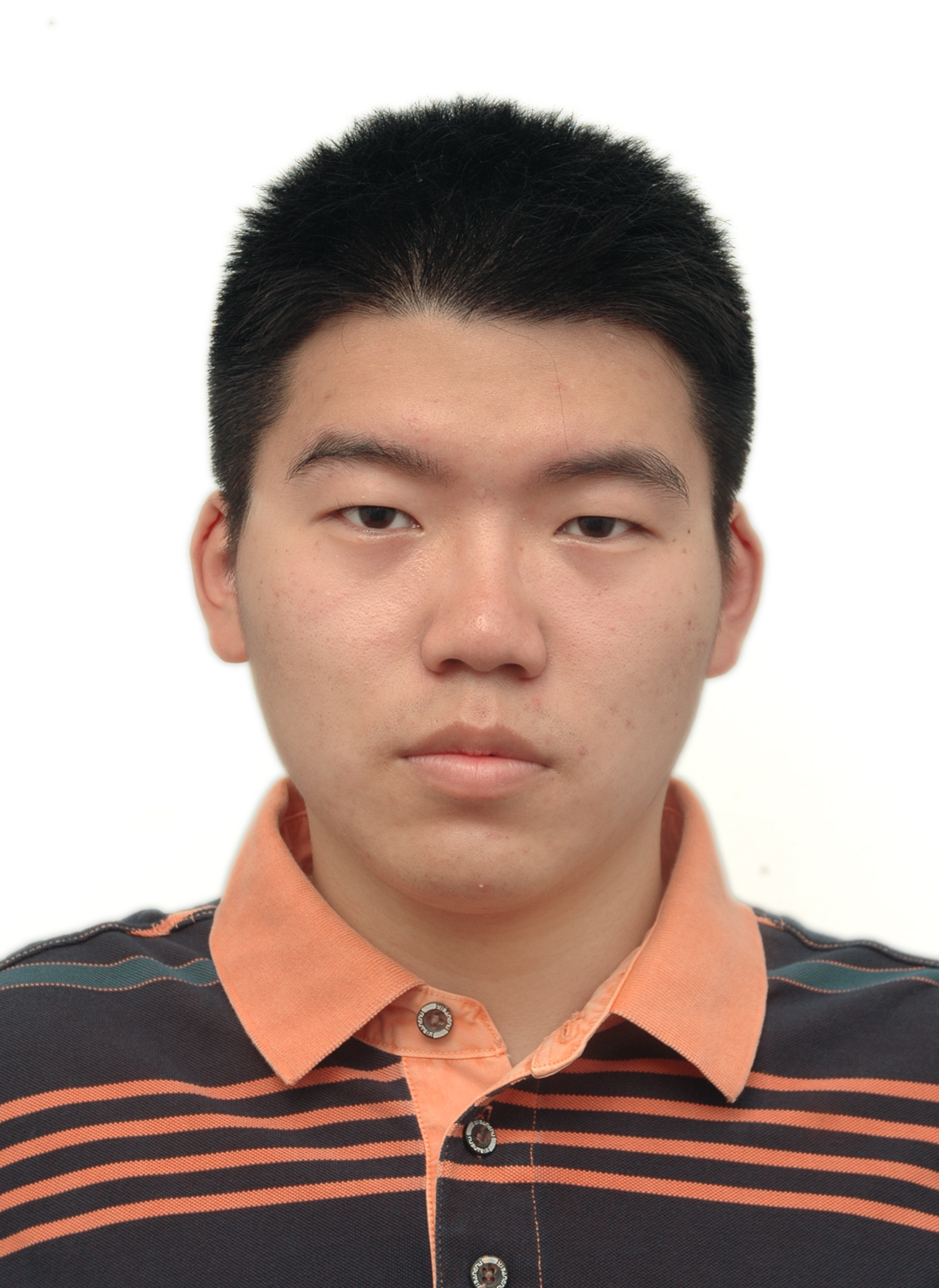}}]{Yang Huang}
Yang Huang received the B.S. and M.S. degrees from Northeastern University, China, in 2011 and 2013, respectively, and the Ph.D. degree from Imperial College London in 2017. He is now an Associate Professor in the Department of Information and Communication Engineering, Nanjing University of Aeronautics and Astronautics, Nanjing, China. His research interests include wireless communications, MIMO systems, convex optimization, machine learning, signal processing for communications, 5G networks, the Internet of Things, wireless power transfer and wireless powered communications.
\end{IEEEbiography}

\vfill

\newpage

\begin{IEEEbiography}[{\includegraphics[width=1in,height=1.25in,clip,keepaspectratio]{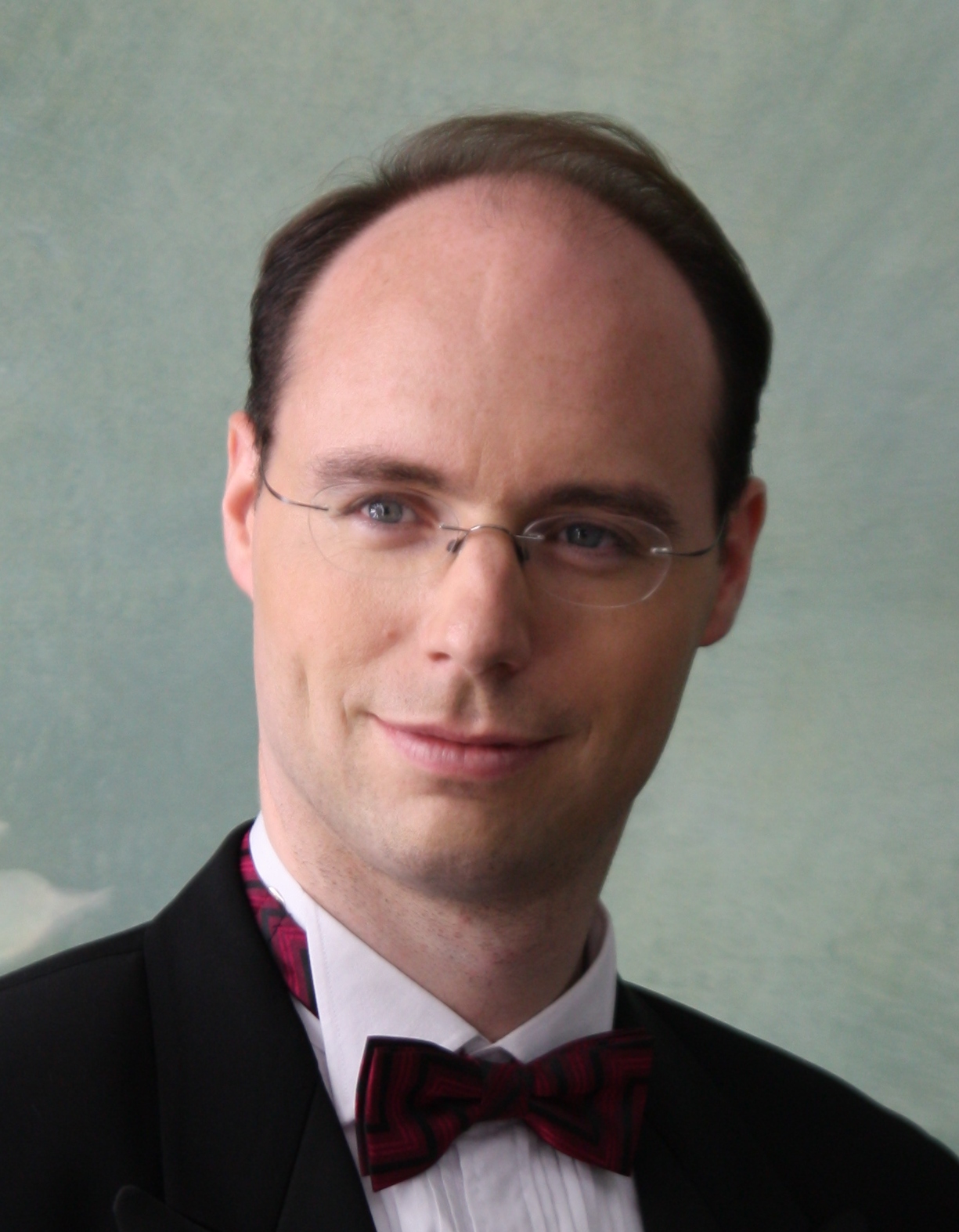}}]{Bruno Clerckx}
Bruno Clerckx (SM’17) received the M.S. and Ph.D. degrees in applied science from the Université Catholique de Louvain, Louvain-la-Neuve, Belgium, in 2000 and 2005, respectively. From 2006 to 2011, he was with Samsung Electronics, Suwon, South Korea, where he actively contributed to 3GPP LTE/LTE-A and IEEE 802.16m and acted as the Rapporteur for the 3GPP Coordinated Multi-Point (CoMP) Study Item. From 2014 to 2016, he was an Associate Professor with Korea University, Seoul, South Korea. He also held visiting research appointments at Stanford University, EURECOM, the National University of Singapore, and The University of Hong Kong. Since 2011, he has been with Imperial College London, first as a Lecturer from 2011 to 2015, then as a Senior Lecturer from 2015 to 2017, and now as a Reader. He is currently a Reader (Associate Professor) with the Electrical and Electronic Engineering Department, Imperial College London, London, U.K.

He has authored two books, 150 peer-reviewed international research papers, and 150 standards contributions, and is the inventor of 75 issued or pending patents among which 15 have been adopted in the specifications of 4G (3GPP LTE/LTE-A and IEEE 802.16m) standards. His research area is communication theory and signal processing for wireless networks. He has been a TPC member, a symposium chair, or a TPC chair of many symposia on communication theory, signal processing for communication and wireless communication for several leading international IEEE conferences. He is an Elected Member of the IEEE Signal Processing Society SPCOM Technical Committee. He served as an Editor for the IEEE TRANSACTIONS ON COMMUNICATIONS from 2011 to 2015 and is currently an Editor for the IEEE TRANSACTIONS ON WIRELESS COMMUNICATIONS and the IEEE TRANSACTIONS ON SIGNAL PROCESSING. He has also been a (lead) guest editor for special issues of the EURASIP Journal on Wireless Communications and Networking, IEEE ACCESS and the IEEE JOURNAL ON SELECTED AREAS IN COMMUNICATIONS. He was an Editor for the 3GPP LTE-Advanced Standard Technical Report on CoMP.
\end{IEEEbiography}
\end{document}